\documentclass[a4paper,11pt]{article}
\pdfoutput=1 

\usepackage{jcappub} 

\usepackage[T1]{fontenc} 

\usepackage{newtxtext}
\usepackage{adjustbox}

\usepackage[T1]{fontenc}

\usepackage{lipsum}
\usepackage{mathrsfs}
\usepackage{color}
\usepackage{mathptmx}
\usepackage[T1]{fontenc}
\usepackage{ae,aecompl}
\usepackage{svg}
\usepackage{graphicx}	
\usepackage{amsmath}	
\usepackage{amssymb}	
\usepackage{mathptmx}

\newcommand{\MUV}{\mathrm{M_{UV}}}
\newcommand{\LUV}{\mathrm{L_{UV}}}

\newcommand{\lya}{\mathrm{{Ly\alpha}}}

\newcommand{\fescC}{f_\mathrm{{LyC,esc}}}

\newcommand{\Nion}{\mathrm{\dot{N}_{ion}}}

\newcommand{\comment}[1]{}

\newcommand{\GammaHI}{$\Gamma_\mathrm{HI}$}

\title{On the correlation between Ly$\alpha$ forest opacity and galaxy density in late reionization models }

\author[a]{Nakul Gangolli,}
\author[a]{Anson D'Aloisio,}
\author[b]{Christopher Cain,}
\author[a]{George D. Becker,}
\author[a]{and Holly Christenson}

\affiliation[a]{Department of Physics and Astronomy, University of California, Riverside, \\ 900 University Ave., Riverside, CA, 92521, USA\\}
\affiliation[b]{School of Earth and Space exploration, Arizona State University\\ Tempe, AZ 85281, USA}

\emailAdd{ngang002@ucr.edu}
\emailAdd{ansond@ucr.edu}
\emailAdd{clcain3@asu.edu}
\emailAdd{georgeb@ucr.edu }
\emailAdd{hollychristenson@gmail.com}

\abstract{Observations of quasar absorption spectra provide strong evidence that reionization extended below $z=6$. The relationship between Ly$\alpha$ forest opacity and local galaxy density (the opacity-density relation) is a key observational test of this scenario. Using narrow-band surveys of $z\approx 5.7$ Ly$\alpha$ emitters (LAEs) centered on quasar sight lines, Ref. \cite{Christenson2023} showed that two of the most transmissive Ly$\alpha$ forest segments at this redshift intersect under-densities in the galaxy distribution.  This result is in tension with models of a strongly fluctuating ionizing background, including some models of late reionization, which predict that the vast majority of these segments should intersect over-densities where the ionizing intensity is strongest. In this paper, we use radiative transfer simulations to explore in more detail the opacity-density relation predicted by late reionization models. We find that fields like the one toward quasar PSO J359-06 -- the more under-dense of the two transmissive sight lines in Ref. \cite{Christenson2023} -- are typically associated with recently reionized gas inside of cosmic voids where the hotter temperatures and rarefied densities enhance Ly$\alpha$ transmission. The opacity-density relation's transmissive end is sensitive to the amount of neutral gas in the voids, as well as its morphology, set by the clustering of reionization sources.  These effects are, however, largely degenerate with each other.  We demonstrate that models with very different source clustering can nonetheless yield nearly identical opacity-density relations when their reionization histories are calibrated to match Ly$\alpha$ forest mean flux measurements at $z<6$. In models with fixed source clustering, a lower neutral fraction increases the likelihood of intersecting hot, recently reionized gas in the voids, increasing the likelihood of observing fields like PSO J359-06.  For instance, the probability of observing this field is $15\%$ in a model with neutral fraction $x_{\rm HI} = 5\%$ at $z=5.7$, three times more likely than in a model with $x_{\rm HI} =  15\%$. The opacity-density relation may thus provide a complementary probe of reionization's tail end.}

\keywords{intergalactic media; reionization; high redshift galaxies}

\begin{document}
\maketitle
\flushbottom

\section{Introduction} \label{sec:intro}
    
Quasar absorption spectra measurements have played a prominent role in sharpening our understanding of the reionization process.  One such measurement is the large scatter in the Ly$\alpha$ forest effective optical depth, $\tau^{50}_{\rm eff}$, at $z\gtrsim 5.3$ \citep{Fan2006, Becker2015, Eilers18, Bosman18, Bosman2021, Yang2020}.  At these redshifts, the extended high-opacity tail of the $\tau^{50}_{\rm eff}$ distribution consists of long absorption troughs, epitomized by the extreme 110 $h^{-1}$cMpc-long trough in the sight light of quasar ULAS J0148+0600 \cite{Becker2015} (J0148 hereafter). The scatter in $\tau_{\rm eff}$ and the existence of such long troughs are inconsistent with standard models of the low-redshift intergalactic medium (IGM), which adopt a uniform ionizing background (as an approximation) and a relaxed temperature-density relation -- characteristics expected well after the end of reionization \cite{Bosman2021}. 

Early ideas to account for the $\tau^{50}_{\rm eff}$ scatter included relic temperature fluctuations from reionization \citep{DAloisio2015} and large ionizing intensity fluctuations from a spatially fluctuating mean free path \citep{Davies2016} and/or rare sources such as active galactic nuclei \citep{2017MNRAS.465.3429C}. In the relic temperature fluctuation scenario, the most opaque regions of the $z=5-6$ forest would be the cosmological density peaks that were reionized at early times, which have had a longer time to cool to lower temperatures. The most transmissive regions would be hot voids that were ionized and heated at the end of reionization. On the other hand, the intensity fluctuation scenarios generally predict the opposite; the density peaks would be the most transmissive, since the ionizing sources are clustered there, while the source-poor voids would be the most opaque.\footnote{The degree of correlation between cosmological density and forest transmission depends on the clustering of the sources. (Although, we will show here that the effect of clustering is degenerate with the reionization history.) The shot noise dominated AGN scenario of Ref. \cite{Chardin2017} can result in less correlation between the forest transmission and environmental density \cite{Davies2018a}.}  

These opposing predictions led Ref. \cite{Davies2018a} to propose large-scale structure surveys centered on Ly$\alpha$ forest sight lines as a test of the proposed models. They showed that the models could be tested by measuring the correlation between the Ly$\alpha$ forest $\tau^{50}_{\rm eff}$ and the central galaxy surface density (within an impact parameter of $10~h^{-1}$cMpc). In what follows, we will refer to this as the opacity-density relation.  The first such survey was presented in Ref. \cite{Becker2018}. They mapped the distribution of strong Ly$\alpha$ emitting galaxies (LAEs) around the extreme Ly$\alpha$ trough toward J0148 and showed that it intersects a large-scale void, therefore disfavoring the possibility that the $\tau^{50}_{\rm eff}$ scatter is driven by relic temperature fluctuations alone. Subsequent studies confirmed this conclusion using Lyman Break Galaxies (LBGs) \citep{2020ApJ...888....6K}, as well as a different field toward another long trough in the forest of quasar SDSS J1250+3130 (J1250 hereafter) \cite{Christenson2021}.   

Soon after the measurement of Ref. \cite{Becker2018}, it was realized that the large scatter in $\tau^{50}_{\rm eff}$, as well as the connection between highly opaque forest segments and cosmic under-densities, could be explained more naturally by a model in which reionization ends at $z < 6$ \cite{Kulkarni19, Keating20, NasirDAloisio19}.  As pointed out by Ref. \cite{Kulkarni19}, extending the tail end of reionization below $z=6$ appears necessary for models to latch onto the observed evolution of the Ly$\alpha$ forest mean transmission. The model has also received support from measurements of dark gap statistics in the coeval Ly$\alpha$ and Ly$\beta$ forests \cite{Zhu2021,Zhu2022}, and the mean free path of ionizing photons between $z=5-6$ \cite{Becker2021, 2021arXiv210812446B, Zhu2023}. In the late reionization scenario, the voids contain isolated patches of neutral gas where reionization was still ongoing (termed ``neutral islands"), which generate the long forest troughs \cite{2015ApJ...799..179M}.  Recently, Ref. \cite{2024arXiv240508885B} reported strong evidence for damping wing absorption at the boundary of the long trough toward J0148.  Refs. \cite{2024MNRAS.tmpL..59Z,2024arXiv240512273S} also found evidence for damping wings in stacks of long dark gaps at $z\gtrsim 5.5$.  These findings add significantly to the evidence that the long opaque segments of the forest at these redshifts are caused by neutral islands. 

As a consensus was forming around the late reionization model, Ref. \cite[C23]{Christenson2023} conducted large-scale structure surveys around 2 of the most transmissive (low $\tau^{50}_{\rm eff}$) forest sight lines toward quasars SDSS J1306+0356 and PSO J359-06 (J1306 and J359), and an additional 3 quasar sight lines were surveyed in Ref. \cite{Ishimoto2022}. With the acquisition of these surveys, it is now possible to make a rudimentary comparison between models and measurements of the opacity-density relation. Interestingly, C23 found that the fields toward J1306 and J359 are both under-dense in LAEs within a radius of $10~h^{-1}$ cMpc of their respective forest sight lines. At face value, this finding is in tension with the intensity fluctuation model of Ref. \cite{Davies2016} as well as the late reionization models of Ref. \cite{NasirDAloisio19}, which predict similarly strong anti-correlations between $\tau^{50}_{\rm eff}$ and galaxy density.  

That the models of Refs. \cite{NasirDAloisio19} and \cite{Davies2016} yield similar predictions should not come as a surprise, however. In Ref. \cite{NasirDAloisio19}, the ionizing backgrounds in their late reionization simulations are modeled using similar techniques and approximations to radiative transfer as those in \cite{Davies2016}.  A chief goal of the current paper is to examine in more detail whether the C23 results are in tension with the late reionization scenario using full radiative transfer simulations of reionization. To this end, we will use several models for the sources of reionization. This allows us to address the question of whether assumptions about the sources are responsible for the tension in C23.  More broadly, the opacity-density relation links the galaxies to the Ly$\alpha$ transmission properties of their environments. Another natural question, then, is whether the relation could tell us which galaxies sourced the high-$z$ ionizing background. For instance, it has been recently proposed that the observed UV-bright LAEs were the sources of reionization, since, it is argued, they should have been prodigious leakers of ionizing radiation \cite[e.g.][]{Naidu19, Mathee2022v1} (see however \cite{2024arXiv240607618C}). Could a measurement of the LAE opacity-density relation test this proposal?   Lastly, we will use our simulations to interpret the fields of J1306 and J359 within the context of late reionization models.  We will discuss the physical conditions of the IGM that could give rise to these fields.  

The remainder of this paper is outlined as follows. In \S \ref{sec:Methods} we summarize our simulation methods.  In \S \ref{sec:em_models} we describe our reionization models. In \S \ref{sec:results} and \S \ref{sec:reion_timing} we present our main results.  In \S \ref{sec:conclusion} we conclude.  In what follows, we adopt a standard flat $\Lambda$CDM cosmology with $\Omega_{M} = 0.305$, $\Omega_{\Lambda} = 1-\Omega_{M}$, $\Omega_{b} = 0.048$, $H_0 = 100 h$ km/s/Mpc, with $h=0.68$. Our simulations are initialized with a linear matter power spectrum with $\sigma_8 = 0.82$ and $n_s = 0.9667$. From here on, all distances are quoted in comoving units. We sometimes use cMpc to denote comoving Mpc.  

\section{Methodology} \label{sec:Methods}

\subsection{Radiative transfer simulations of reionization}
\label{subsec:rt_sim}

We ran multi-frequency radiative transfer (RT) simulations of reionization using the \textsc{FlexRT} code of Ref. \cite{2024JCAP...12..025C} (see also\cite{Cain2021,Cain2022,Cain2023}).  Here we summarize the essential details for this paper. \textsc{FlexRT} is run in post-processing on density fields extracted from a cosmological simulation. The density fields used here are from the hydrodynamics simulation of~\cite{DAloisio2018}, with a box size of $L=200$ $h^{-1}$cMpc. The original simulation was run with the Eulerian hydrodynamics code of Ref. \cite{2004NewA....9..443T} using $N = 2\times 2048^3$ dark matter particles and gas resolution elements. A key improvement for the current paper is that we ran a higher resolution N-body-only simulation using the same initial conditions with $N=3600^3$ particles, allowing us to resolve halos of smaller masses.  Hence, we no longer use the sub-resolution halo algorithm use in \cite{Cain2021, Cain2022}.\footnote{Their sub-resolution halo algorithm was adapted from the non-linear bias method of Ref. \cite{2015MNRAS.450.1486A}.}  As stated in Ref. \cite{Cain2023}, our new simulation is complete down to a mass limit of $M_{\rm min}=3 \times 10^{9}~h^{-1} \mathrm{M}_{\odot}$ at $z=5.7$ . (Here we define our simulated mass function to be complete if it agrees to within 10\% with the mass function of Ref. \cite{Trac_2015}, which was constructed from a suite of higher resolution simulations.)  The schemes that we explore for assigning ionizing photon emissivities to halos are described in \S\ref{sec:em_models}.
 
The RT is performed on a uniform grid with $N_{\rm rt} = 200^3$ cells.  We adopt a source spectrum of $J_{\nu} \propto \nu^{-1.5}$ between 1 and 4 Ry, discretized in 5 frequency bins centered on [14.48,16.70,20.03,25.78,39.23] eV.  The bins are chosen to contain an initially equal fraction of photons for the assumed source spectrum. \textsc{FlexRT} is based on the adaptive ray tracing method of Ref. 
 \cite{2002MNRAS.330L..53A} (see also Ref. \cite{2007ApJ...671....1T}).  Rays are cast from the centers of source cells and can adaptively split and merge to maintain a user-specified angular resolution in the radiation field.  For the simulations in this work, rays were cast and merged at HealPix level $l=0$, corresponding to 12 directions per source cell.  Merging begins for rays above a minimum number of 16 rays per cell. Generally speaking, a lower number of rays yields a noisier radiation field. In Appendix \ref{app:ray_noise}, we compare the ionizing background in our fiducial simulation to a simulation with higher angular resolution run at HealPix level $l=1$ (48 directions).  We also compare to an averaging procedure that smooths over the noise.  We demonstrate that our main results are not significantly affected by ray noise.  

The coarse-grained cosmological density fields for the RT are obtained by averaging with equal weights over the hydrodynamic simulation cells within a given RT cell. The RT cells are $\Delta x = 1~h^{-1}$cMpc wide; too large to resolve directly the small-scale gas structures that set the LyC opacity. To remedy this, we apply the sub-grid model of Ref. \cite{Cain2021} for the IGM opacity.  This model is based on an extended suite of the high-resolution radiative hydrodynamics simulations of Refs. \cite{DAloisio2020,Nasir2021}, which track the evolution of the LyC opacity as the IGM responds to photoheating from reionization.  The simulations upon which the sub-grid model is constructed have boxes sizes with $L_{\rm box} = 1~h^{-1}$Mpc, equal to an RT cell size.  The suite spans a range of environmental parameters --local photo-ionization rate ($\Gamma_{
\rm HI}$), box-scale density, and reionization redshift ($z_{\rm reion}$) -- to capture variations in the opacity owing to the patchiness of reionization. For each RT cell, these quantities are mapped to an opacity evolution in a way that accounts for the photoionization history of the cell (see Refs. \cite{Cain2021,Cain2022} for more details). 

We use the full speed of light because the photoionization rate of the IGM near the end of reionization is highly sensitive to the reduced speed of light approximation \citep{Ocvirk2019,Cain2023} (see also Cain in prep.). \textsc{FlexRT} tracks the photoionization and thermal states of each RT cell. For the latter, we use the fitting function of Ref. \cite{DAloisio2019} for the peak temperatures achieved behind ionization fronts ($T_{\rm reion}$), which was calibrated to a suite of high-resolution one-dimensional RT simulations. The equilibrium photoheating in the highly ionized IGM includes spectral filtering as described in Ref. \cite{Cain2023}.    

\subsection{The Ly$\alpha$ Forest}
\label{subsec:lyaforest}

To model the Ly$\alpha$ forest, we trace 4,000 skewers in random directions and starting at random locations in the periodic box of our hydrodynamic simulation. We compute the Ly$\alpha$ transmission along the skewers using the approximation to the Voigt profile of Ref. \cite{2006MNRAS.369.2025T}.  Applying the hydrogen photoionization rates ($\Gamma_{\rm HI}$) and temperatures ($T$) from the RT grid, the neutral hydrogen densities in reionized regions along the hydro skewers are rescaled under the assumption of photoionization equilibrium with the case-A recombination coefficient.\footnote{Photoionization equilibrium and the case-A recombination coefficient are good approximations in the highly ionized $\Delta \lesssim 1$ regions that set the forest transmission at these redshifts.}  RT cells with ionized fraction $x_{\rm ion} < 0.5$ are taken to be fully neutral, but we find that the $\tau^{50}_{\rm eff}$ values of our mock Ly$\alpha$ forest are not sensitive to the particular choice of the $x_{\rm ion}$ threshold. 

We note that the resolution of our hydro simulation ($N_{\rm gas}=2048^3$) is much higher than that of our RT grid ($N_{\rm RT} = 200^3$). One issue with the approach outlined so far is that is misses the effects of temperature variations on scales smaller than our coarse-grained RT cells of $\Delta x = 1~h^{-1}$cMpc.  This is less of a problem shortly after a cell has been reionized, when it is approximately isothermal with temperature $\approx T_{\rm reion}$. Well afterwards, however, hydro cells with lower (higher) densities than the local RT-cell-averaged mean should have lower (higher) temperatures than the value assigned to the RT cell.  This leads to an over-estimate of the temperatures in small-scale under-densities which, in turn, over-estimates the forest transmission by $\approx 10-15\%$.  

We correct for this effect by assigning a local temperature-density relation to each RT cell as in \cite{Cain2023}.  Using the analytic solution of the temperature evolution equation from Ref. \cite{McQuinn2016}, we evaluate the IGM temperature on a grid of ($z$, $z_{\rm reion}$, $\Delta$) values, where $\Delta$ is gas density in units of the cosmic mean.  For each $z$ and $z_{\rm reion}$, we evaluate $\gamma$ by fitting densities with $\Delta < 1$ (the regime that sets the forest transmission) to a power law.  We then assign a local value of $\gamma$ to each RT cell by interpolating in $z$ and $z_{\rm reion}$.  The temperature of each hydro cell is then given by
\begin{equation}
    T_{\rm hydro} = T_{\rm RT}\left(\frac{\Delta_{\rm hydro}}{\Delta_{\rm RT}}\right)^{\gamma(z,z_{\rm reion}) - 1},
\end{equation}
where $T_{\rm RT}$ and $\Delta_{\rm RT}$ are the temperature and density of the coarse-grained RT cell, $\Delta_{\rm hydro}$ is the density along the hydro skewer intesecting the RT cell, and $\gamma(z,z_{\rm reion})$ represents the slope of the local temperature-density relation. We refer the reader to Appendix E of Ref. \cite{Cain2023} for a more detailed discussion and testing of this procedure.  

Lastly, it is well known that achieving numerical convergence in simulations of the high-$z$ Ly$\alpha$ forest is difficult \citep[see e.g.][]{2009MNRAS.398L..26B, DAloisio2018}. The appendix of Ref. \cite{DAloisio2018} describes a convergence test for the same hydrodynamic simulation used in this work.  We apply the correction factors provided in their Table A1 to the neutral hydrogen densities in reionized regions to correct for resolution in our Ly$\alpha$ forest fluxes.  The correction increases the transmitted flux in our simulated forest, accounting for the fact that small-scale voids are deeper in more highly resolved simulations. For a fixed mean Ly$\alpha$ forest flux, this requires our simulations to end reionization somewhat later compared to a simulation without the correction (see \S 6.2 of \cite{Cain2023}). We will return to point below when we discuss how the timing of reionization affects the opacity-density relation. 

\subsection{Ly$\alpha$ Emitters}
\label{subsec:LAEs}

To populate our halos with LAEs, we follow the approach outlined in Refs. \cite{Weinberger2019,Gangolli2021}. In summary, we first assign rest-frame UV magnitudes to the halos by abundance matching to the luminosity function of Ref. \cite{Finklestein2019}. For a given UV magnitude, we then draw a $\lya$ rest-frame equivalent width from the empirically calibrated distribution of Ref. \cite{Dijkstra2012}.   

We generate mock LAE surveys as described in Ref. \cite{NasirDAloisio19}.  We first select LAEs within cylinders of radii $100~h^{-1}$cMpc and depths $60~h^{-1}$cMpc, centered on each of the 4,000 Ly$\alpha$ forest sight lines traced through our box. (In the end, LAEs are selected by magnitude and color cuts as described below; here the depth of $60~h^{-1}$cMpc ensures that the preliminary selection spans the width of the filter.)  Following Refs. \cite{Keating20,NasirDAloisio19}, we model the continuum as a power law, $F_{\lambda} \propto \lambda^{-2}$, and the Ly$\alpha$ line as two Gaussian peaks blue-ward and red-ward of systemic, with the former completely attenuated.  The red-ward peak is characterized by two parameters: the velocity offset ($v_{\mathrm{off}}$) and the line width ($\sigma_{v}$).  We map the UV magnitude of a galaxy to a $v_{\mathrm{off}}$ value using the scaling relation in Equation 3 of Ref. \cite{Mason2018v1}.  For simplicity, we follow Ref. \cite{Mason2018v1} in setting $\sigma_{v} = v_{\mathrm{off}}$, reflecting the expectation that a line's width is correlated with its offset.   Skewers are then traced along the line of sight from each LAE.  The spectrum of a given LAE is attenuated according to the gas properties along the corresponding skewer, derived from the hydro and \textsc{FlexRT} simulations, with the rest-frame defined as the halo's frame.\footnote{ In computing the transmitted fraction for each LAE spectrum, we use a bin size of $\Delta v = 3.5$ km/s, linearly interpolating between our hydro cells (which have a native $\Delta v = 14$ km/s). We find that the interpolation to higher resolution is necessary to avoid spurious transmission spikes in regions with large velocity gradients, owing to the lower native resolution of our hydro skewers. We have tested for convergence of the transmitted fraction with respect to $\Delta v$.} We note that our hydro simulation does not include prescriptions for star-formation and feedback processes, and the resolution is not high enough to model the interstellar and circumgalactic gases which shape the emerging Ly$\alpha$ line.   Instead, the approach taken here is to roll their effects into our parameterization of the line profile assigned to each halo, i.e. we assume that the red-side Gaussian is the emerging line profile. To avoid ``double counting'' attenuation from the gas close to the halo, we skip the first 3 hydro cells of each skewer, which corresponds to a comoving distance of $0.3~h^{-1}$cMpc.

After attenuation, we calculate the NB816 and $i2$ magnitudes of the LAEs using publicly available transmission curves of these filters on the Subaru Hyper Suprime-Cam.  Finally, we apply magnitude and color cuts of NB816 $\leq$ 25.5 and $i2-$NB816 $\geq 1.2$, the same as those applied in C23. C23 defined the mean surface density of a given LAE field, $\langle \Sigma_{\rm LAE} \rangle$, to be the surface density measured within an annulus of inner and outer radii 15 and 40 arcminutes, respectively. We will do the same here.  The mean value of $\langle \Sigma_{\rm LAE} \rangle$ among the 4 fields in C23 is $0.022~(h^{-1} \mathrm{cMpc})^{-2}$.  (We emphasize that the $\langle \ldots \rangle$ denotes the average value in one field as defined above.) The LAE fields in all of our models are calibrated to match approximately this mean value. To achieve this, we perturb $v_{\mathrm{off}}$ by a constant value for all fields in a given model.  After these adjustments, the average values of $\langle \Sigma_{\rm LAE} \rangle$ are in the range $0.021-0.022$ ($h^{-1}$ Mpc)$^{-2}$ in all of the models presented below. 

\section{Reionization Models} \label{sec:em_models}

In this section we describe our source models (\S \ref{sec:source_models}) and how they are calibrated to satisfy the empirical boundary conditions for reionization provided by quasar absorption spectra measurements (\S \ref{sec:lyaforest_calib}).

\subsection{Source models}
\label{sec:source_models}

In what follows we will compare results from the following source models.

\begin{figure}
\begin{center}
\resizebox{0.9\columnwidth}{!}{\includegraphics{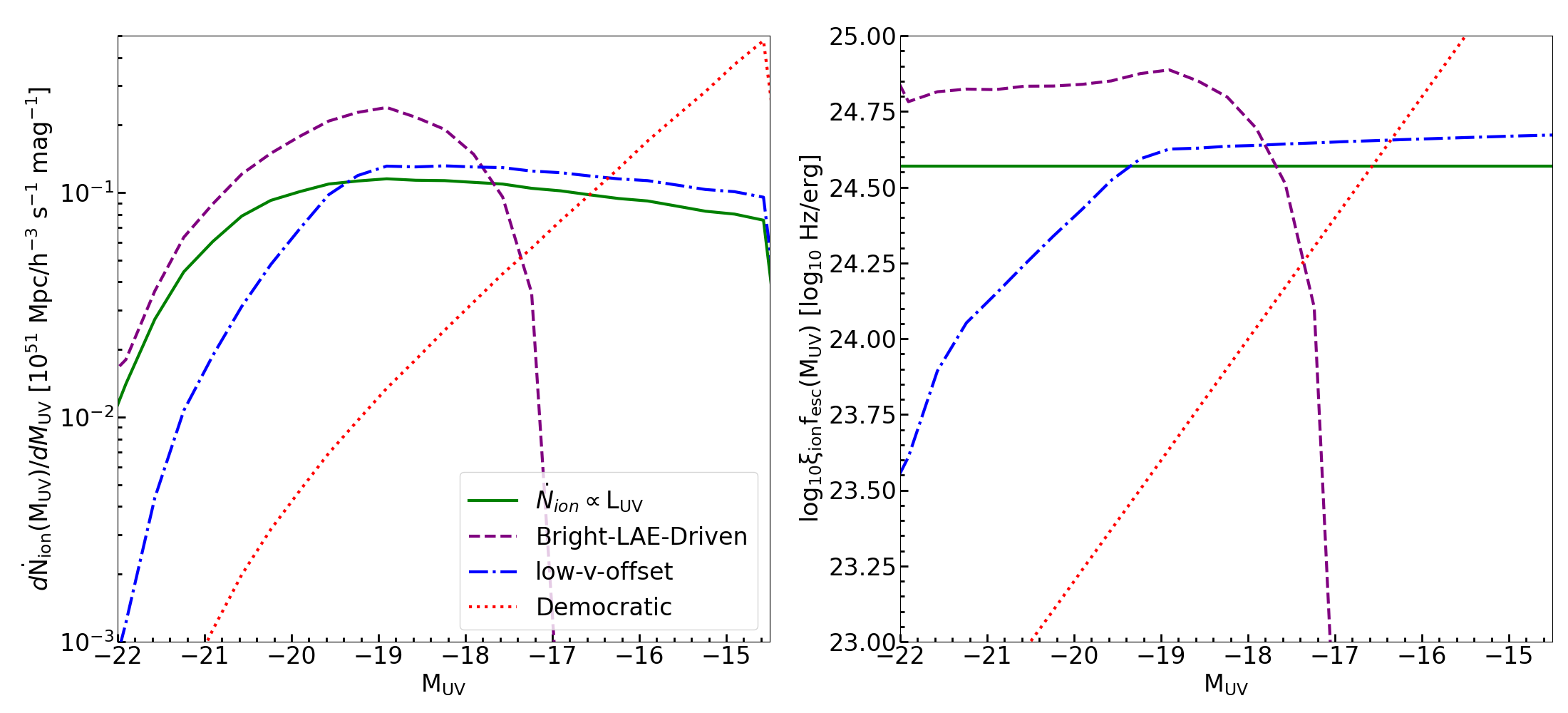}}
\end{center}
\caption{Ionizing photon output at $z=5.7$ in our four source models. {\it Left:} Contribution to the ionizing emissivity per unit UV magnitude. {\it Right:} product of effective ionizing photon production efficiency and LyC escape fraction, $\xi_{\mathrm{ion}} f_{\mathrm{esc}}$, defined to be the ratio of the ionizing emissivity to the UV luminosity density.  The models span a wide range of possibilities for the sources of reionization.  For instance, in the \textsc{bright-LAE-driven} model, the observed UV-bright LAEs are the sole leakers of ionizing radiation. In the \textsc{democratic} model, the sources are predominantly the faintest galaxies in the UV luminosity function, well below current detection limits.  We use these models to explore how different assumptions about reionization's sources affect the opacity-density relation.}
\label{fig:ion_emissivity}
\end{figure}

\begin{itemize}

\item {\textsc{$\dot{N}_{\mathrm{ion}} \propto L_{\rm UV}$}} (green): This model adopts the standard assumption that the rate of ionizing photon production of a galaxy is proportional to its UV luminosity, $\dot{N}_{\mathrm{ion}} \propto L_{\rm UV}$.  The constant of proportionality relating $L_{{\rm UV}}$ to $\dot{n}_{\rm ion}$ is a redshift-dependent free parameter that we tune to achieve agreement with quasar absorption spectra measurements at $z = 4.8 - 6$ (primarily the mean Ly$\alpha$ forest flux), as we show in the next section. The left panel of Figure \ref{fig:ion_emissivity} shows how the ionizing emissivity varies across the source population for all of our models at $z=5.7$, as a function of absolute UV magnitude, $M_{\rm UV}$. The right panel shows the ratio of the emissivity to the UV luminosity density, which we defined to be $\xi_{\rm ion} f_{\rm esc}$, the product of the effective ionizing photon production efficiency and escape fraction for the model. Notably, $\xi_{\rm ion} f_{\rm esc}$ is independent of the UV magnitude for this model. (We emphasize that $\xi_{\rm ion} f_{\rm esc}$ does evolve with redshift.)   

\item {\textsc{bright-LAE-driven}} (purple): This model assumes that the LyC leakers are the observed UV-bright LAEs.  In contrast to the \textsc{$\dot{N}_{\mathrm{ion}} \propto L_{\rm UV}$} model, the ionizing emissivity of a galaxy is assumed proportional to its Ly$\alpha$ luminosity, $\dot{N}_{\rm ion} \propto L_{{\rm Ly} \alpha}$.  Following Ref. \cite{Mathee2022v2}, we assume that only LAEs with $L_{{\rm Ly} \alpha} \geq 10^{42.2}$ erg/s contribute, and we randomly exclude 50\% of the galaxies in this subset. The latter exclusion is motivated by the findings of Ref. \cite{Naidu2022v1} that roughly half of their sample from the $z \approx 2$ X-SHOOTER survey have a median LyC escape fraction of $\fescC \approx 50\%$, with the other half being non-leakers. Again, the proportionality between $\dot{N}_{\rm ion}$ and $ L_{{\rm Ly} \alpha}$ is a redshift-dependent free parameter that we tune to achieve agreement with $z=4.8-6$ quasar absorption spectra measurements.  

 \item {\textsc{low-v-offset}} (blue):  This model also assumes $\Nion \propto L_{{\rm Ly} \alpha}$. The distinguishing feature is the assumption that the LyC escape fraction is anti-correlated with the Ly$\alpha$ line velocity offset, $v_{\rm off}$.  Again, we use the empirically calibrated $\MUV-v_{\rm off}$ relation given by equation (3) from Ref. \cite{Mason2018v1}, which assigns larger $v_\mathrm{ \rm off}$ to more UV luminous galaxies.  We then map the $v_{\rm off}$ to a LyC escape fraction using equation (2) from Ref. \cite{Izotov2018}.\footnote{As noted in Ref. \cite{Izotov2018}, their relation results in $f_{esc} > 1$ for $v_{off} \lesssim 140$km/s, so we impose the limit $f_{esc} = [0, 1]$.} This prescription associates high LyC leakage with strong Ly$\alpha$ emission at low velocity offsets from systematic. Figure \ref{fig:ion_emissivity} shows that this model contrasts with the \textsc{bright-LAE-driven} model in that many of the ionizing photons come from UV-fainter LAEs.  

 \item{\textsc{democratic}} (red): In this model, we assume that every galaxy contributes the same emissivity, i.e. $\dot{N}_{\mathrm{ion}}$ is independent of $L_{\rm UV}$ and $L_{\mathrm{Ly}\alpha}$. The main purpose of this simplistic model is to provide an extreme example where the faintest galaxies dominate the ionizing photon budget (see Figure \ref{fig:ion_emissivity}), contrasting against the  models above. 
\end{itemize}

The above models are used to test how different assumptions about the sources affect the opacity-density relation.

\subsection{Model Calibration} \label{sec:lyaforest_calib}

\begin{figure*}
\begin{center}
\resizebox{15.cm}{!}{\includegraphics{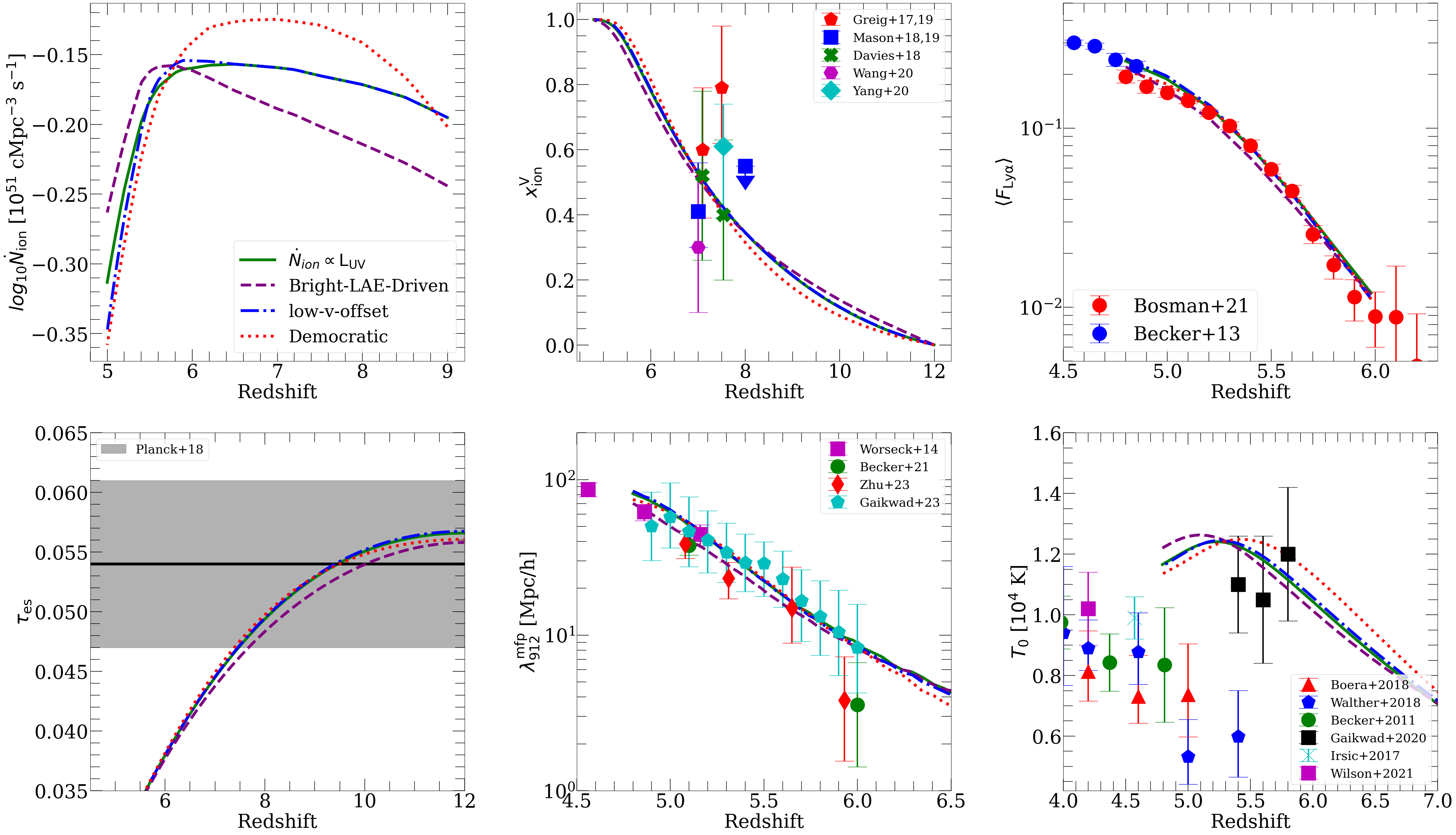}}
\end{center}
\caption{Calibration of our reionization models to match the boundary conditions provided by the Ly$\alpha$ forest.  Starting in the upper left and moving clockwise, we show the: 1) ionizing emissivity; 2) reionization history \cite{Greig2017, Greig2019, Davies18, Wang20, Mason2018v1, Mason2019, Yang2020}; 3) mean fraction of transmitted flux in the Ly$\alpha$ forest \cite{Becker2013, Bosman2021}; 4) IGM temperature at mean density \cite{Becker2011, Walther2018, Boera2019, Gaikwad2020, Wilson2021} ; 5) mean free path at wavelength $912$ \AA \cite{Worseck2014, Becker2021, Gaikwad2023, Zhu2023}; 6) CMB electron scattering optical depth \cite{Planck18}. The ionizing emissivity (top-left) is adjusted as a function of redshift to match the Ly$\alpha$ forest mean flux evolution. Matching the mean flux yields reasonable agreement with the other observables.}
\label{fig:obs_anchors}
\end{figure*}

\begin{figure}
\begin{center}
\resizebox{15.cm}{!}{\includegraphics{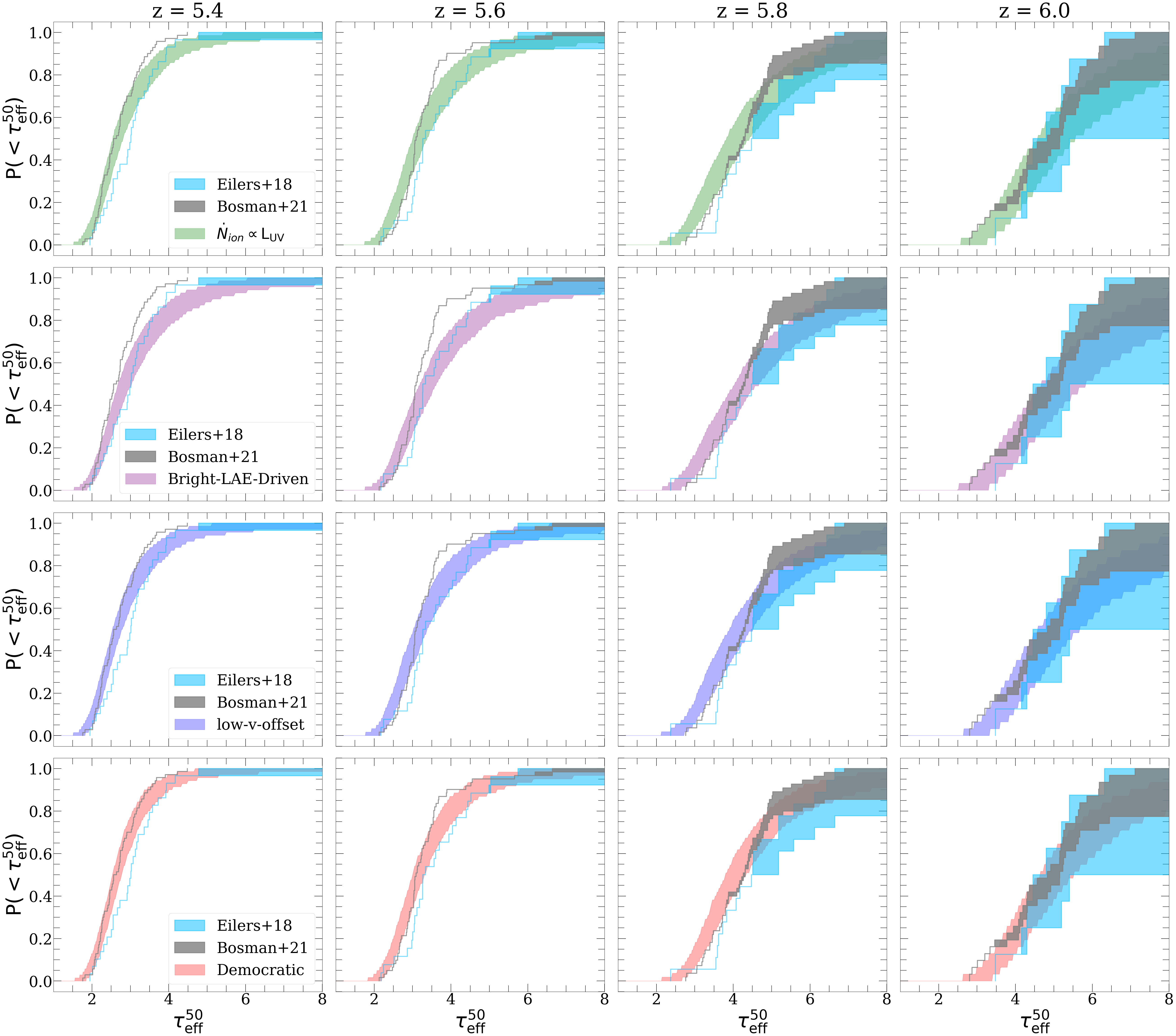}}
\end{center}
\caption{Opacity fluctuations in the Ly$\alpha$ forest.  The panels show the CDFs of the effective optical depth, $\tau^{50}_{\rm eff}$, at redshifts annotated at the top of each panel.  The cyan and gray shading shows the observational measurements of Refs. \cite{Eilers18} and \cite{Bosman2021}.  The model CDFs were obtained by bootstrap sampling using the number of sight lines in each redshift bin from \cite{Bosman2021}, so are most directly comparable to the gray observational results. For each model, the colored shading encompasses 80\% of the simulated CDFs.}
\label{fig:scatter}
\end{figure}

To ensure that each reionization model satisfies the empirical boundary conditions provided by quasar absorption spectra measurements, we have calibrated the ionizing emissivity history, $\dot{N}_{\rm ion}(z)$, for each source model to match the measured evolution of the mean Ly$\alpha$ forest flux from $z=5-6$ \citep{Bosman2021}.  Figure \ref{fig:obs_anchors} shows the results of these calibrations. Starting in the upper left corner, and moving clockwise, we show the: 1) ionizing emissivity; 2) reionization history; 3) mean fraction of transmitted flux in the Ly$\alpha$ forest; 4) IGM temperature at mean density; 5) mean free path at wavelength $912$ \AA; and 6) CMB electron scattering optical depth, $\tau_{es}$.  We find that by calibrating our models to match the mean $\lya$ forest flux, we consequentially recover reasonably well the other observables as well. Notably, in all models the calibration requires a steep drop in the ionizing emissivity below $z=6$.  This is qualitatively consistent with almost all other RT simulations in the literature that similarly latch onto the forest boundary conditions (see Ref. \cite{Cain2023} for a detailed discussion).  

In Figure \ref{fig:scatter}, we compare the cumulative distribution functions (CDFs) of $\tau^{50}_{\mathrm{eff}}$ in our calibrated models against the measured distributions at $z=5.4-6$ from Ref. \cite{Bosman2021}, which are shown as the gray shading. For reference, we also show the measurements of Ref. \cite{Eilers18} in cyan.  The model CDFs were obtained by sampling with replacement $N$ forest lines from our initial set of 4,000 sightlines, 500 times, where $N$ is the number of sight lines in a given redshift bin from the measurements of Ref. \cite{Bosman2021}. The model CDFs are therefore most directly comparable to the gray observational results.  The shaded regions in Figure \ref{fig:scatter} for the models encompass 80\% of the simulated CDFs. Models in which the emissivity is sourced by brighter/rarer galaxies generally exhibit more scatter in $\tau^{50}_{\rm eff}$. This owes largely to the higher neutral fraction in those models. (For reference, the \textsc{$\dot{N}_{\rm ion} \propto L_{\rm UV}$}, \textsc{Bright-LAE-driven}, \textsc{Low-v-offset}, and \textsc{Democratic} models have $z=5.7$ global neutral fractions of $x_{\rm HI} = 0.13$, 0.17, 0.13, and 0.1, respectively.) We note that the model CDFs are somewhat wider than the measurements of Ref. \cite{Bosman2021} across redshift.  This may result from reionization ending slightly later in our simulations than the data suggests.  We will return to this possibility in \S \ref{sec:reion_timing}.  

\section{Results} \label{sec:results} 

\subsection{Opacity-Density Relation} 
\label{sec:opac_dens}

\begin{figure*}
\begin{center}
\resizebox{14.cm}{!}{\includegraphics{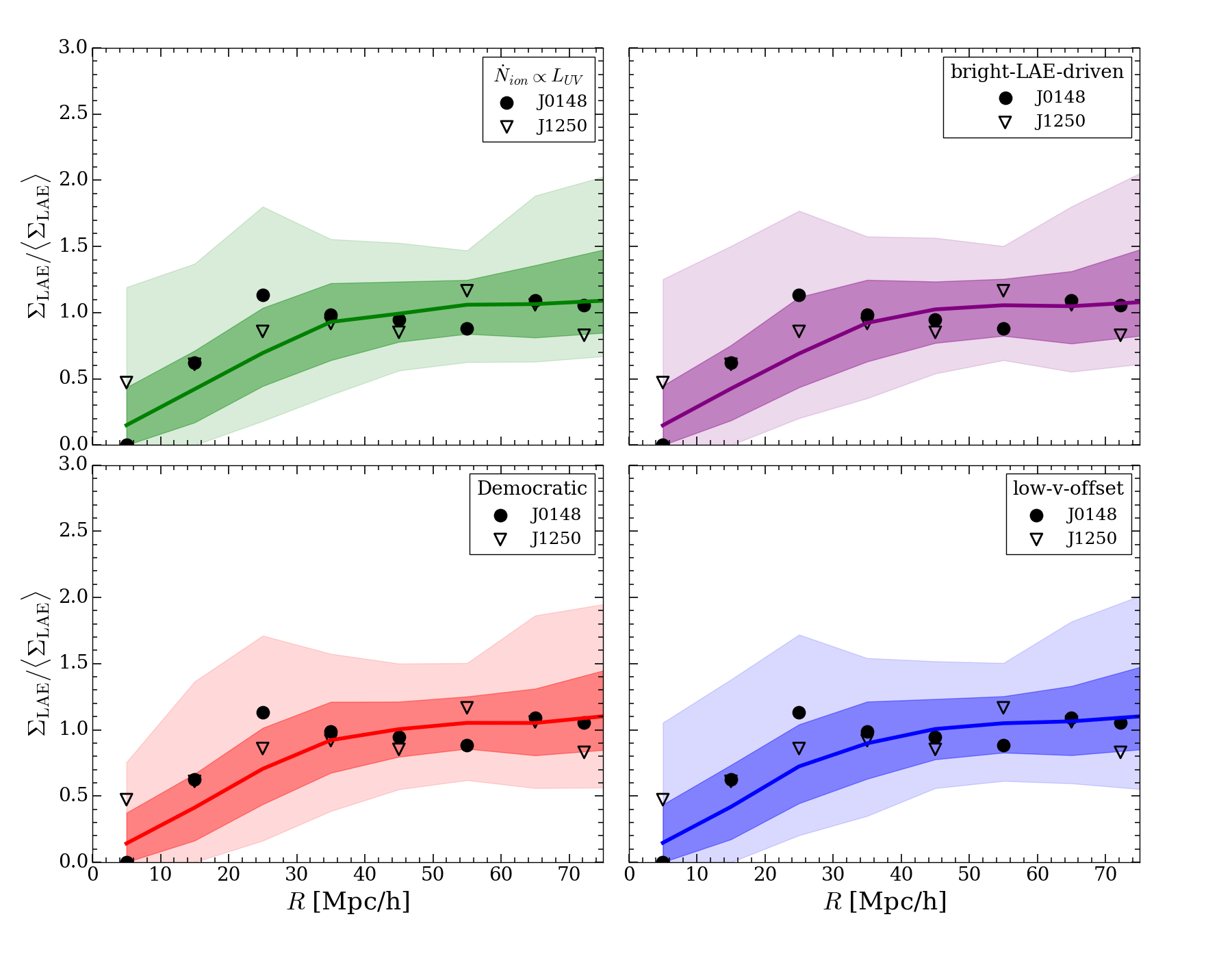}}
\end{center}
\caption{LAE surface density as a function of radius, $R$, from the opaque quasar sight lines ($\tau^{50}_{\rm eff} \geq 7$) in our simulated samples.  The surface densities for each field are normalized by their respective values of $\langle \Sigma_{\rm LAE} \rangle$. The curves show the median profile, while the dark and light shaded regions represent the 68\% and 98\% ranges.  The observational measurements of C23 are shown as filled circles (J0148) and open triangles (J1250). The model results are similar despite their very different assumptions about the ionizing source population. The strong under-density of LAEs near highly opaque forest segments reflects the fact that the voids contain neutral islands, which suppress the forest transmission.}
\label{fig:Sigma_vs_R_opaque}
\end{figure*}

Figure \ref{fig:Sigma_vs_R_opaque} shows the LAE surface density as a function of radius from the opaque quasar sight lines with $\tau^{50}_{\rm eff} \geq 7$. This cut, which yields $200-400$ fields (depending on the model), was chosen to select analogues of the fields toward J0148 and J1250.  Following C23, the surface densities for each field are normalized by their respective values of $\langle \Sigma_{\rm LAE} \rangle$.  The curves show the median surface density profiles, while the dark and light shadings represent the 68\% and 98\% regions.  The filled circles and open triangles show the measurements from J0148 and J1250, respectively, from C23.  As expected, all models exhibit a strong under-density of LAEs toward the center of the field.  This owes to the voids at $z=5.7$ containing neutral islands in all of our reionization models; the most opaque segments of the Ly$\alpha$ forest are those that intersect the voids. The results are highly similar between the models, despite their very different ionizing source populations, highlighting the insensitivity of this statistic to the characteristics of the reionization sources.  

\begin{figure*}
\begin{center}
\resizebox{14.cm}{!}{\includegraphics{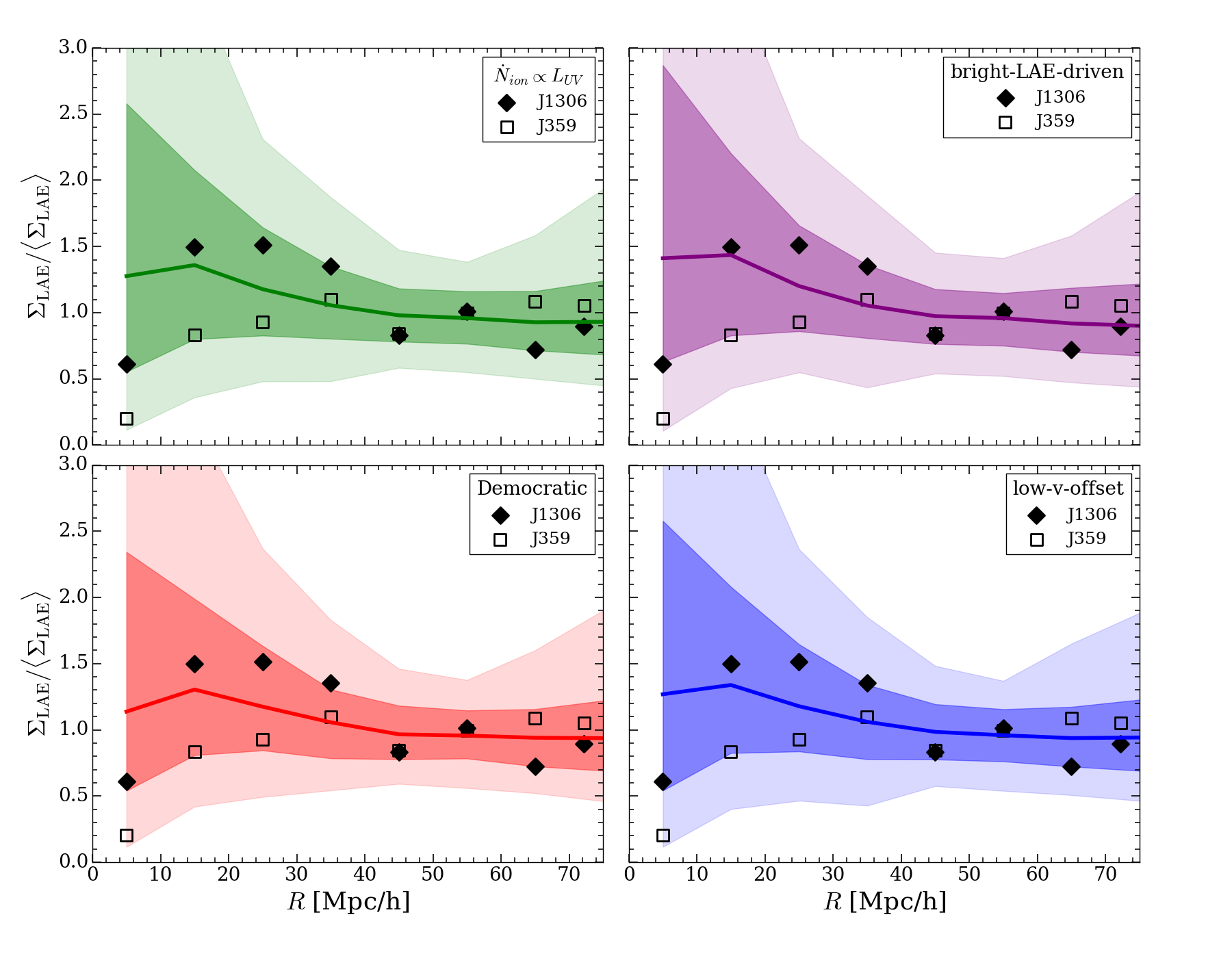}}
\end{center}
\caption{Similar to Figure \ref{fig:Sigma_vs_R_opaque}, except for highly transmissive sight lines ($2.42 \leq \tau_{\mathrm{eff}} \leq 2.92$). Again, the model results are similar despite their very different assumptions about the ionizing source population.}
\label{fig:Sigma_vs_R_transmissive}
\end{figure*}

In Figure \ref{fig:Sigma_vs_R_transmissive} we show the surface density profiles around the most transmissive sight lines, which we define to be those with $2.42 \leq \tau^{50}_{\rm eff} \leq 2.92$. These bounds were chosen to have a central value of $\tau^{50}_{\rm eff}=2.67$, similar to the mean value between J1306 and J359.  The selected sample contains $\sim 600$ fields.  In all models, the fields toward transmissive sight lines typically exhibit mild over-densities, although the scatter is large. The \textsc{bright-LAE-driven} model tends to exhibit stronger LAE over-densities around highly transmissive sight lines. Otherwise, the results are strikingly similar considering the large differences in the source modeling. For instance, in the \textsc{bright-LAE-driven} model, it is the {\it observed} LAEs that provide all of the ionizing photons, whereas in the \textsc{democratic} model the photons come from galaxies well below current detection limits.    

The large scatter in central LAE surface density reflects the fact that the physical conditions leading to high transmission in the forest are varied.  On the one hand, the intense ionizing radiation around galaxy over-densities can lead to enhanced forest transmission. On the other hand, hot, recently reionized regions inside of the voids can be highly transmissive.  More typically the transmission in average regions of the IGM is enhanced by some combination of these physical effects.  Both J1306 and J359 are statistically compatible with the models given the large scatter. Notably, however, fields with the central under-density of J359 are rare in our simulations. (More about the rarity of J359 below.)

\begin{figure*}
\begin{center}
\resizebox{14cm}{!}{\includegraphics{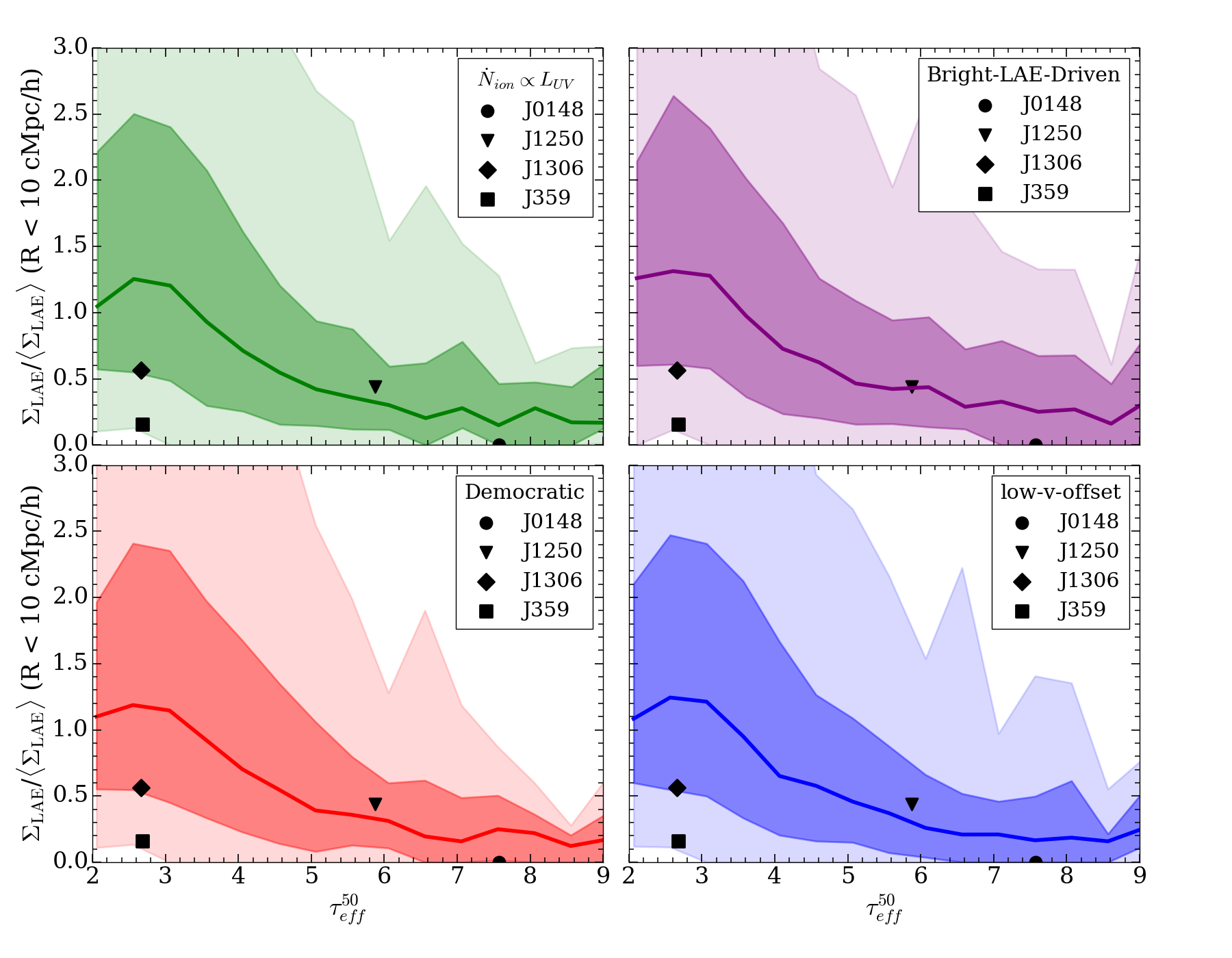}}
\end{center}
\caption{Opacity-density relation in four empirically calibrated reionization models. The curves represents the median central LAE surface density (measured within a circle of radius $10~\mathrm{cMpc}/h$) versus the $\tau^{50}_{\rm eff}$ value of the forest sight line. Dark and light shaded regions represent the 68\% and 98\% ranges, respectively. In the \textsc{Bright-LAE-Driven} model, the highly clustered nature of the sources strengthens the association between low $\tau_{\rm eff}^{50}$ sight lines and galaxy over-densities. Otherwise, the opacity-density relations are quite similar. Strongly under-dense but highly transmissive fields such as the one toward J359 are rare in all of the models, but their probability increases substantially if reionization ends slightly earlier (see \S \ref{sec:reion_timing}).}
\label{FIG:Sigma_vs_taueff}
\end{figure*}

Figure \ref{FIG:Sigma_vs_taueff} compares the opacity-density relations in our models to the four fields in C23. The surface densities are measured within a circular aperture of radius $10~\mathrm{cMpc}/h$.  Again, for each field in the sample, $\Sigma_{\rm LAE}$ is normalized by its corresponding value of $\langle \Sigma_{\rm LAE} \rangle$.  As expected from the previous discussion, there are only modest differences in the opacity-density relations between the models. (We note that the somewhat stronger association between low-$\tau_{\rm eff}^{50}$ sight lines and over-densities in the \textsc{Bright-LAE-Driven} model is evident in the top-right panel).   These results lead us to conclude that, {\it for models that have been calibrated to have the same Ly$\alpha$ forest mean transmission}, the opacity-density relation is broadly insensitive to assumptions about which galaxies sourced the high-$z$ ionizing background. We emphasize, however, that much of this insensitivity owes to: (1) a degeneracy between the morphology of neutral islands (which is set by the source clustering) and the global neutral fraction; (2) the models being calibrated to have the same Ly$\alpha$ forest mean transmission.  As mentioned above, this calibration forces the \textsc{$\dot{N}_{\rm ion} \propto L_{\rm UV}$}, \textsc{Bright-LAE-driven}, \textsc{Low-v-offset}, and \textsc{Democratic} models to have $z=5.7$ global neutral fractions of $x_{\rm HI} = 0.13$, 0.17, 0.13, and 0.1, respectively.  But these differences in $x_{\rm HI}$ largely cancel differences that would otherwise arise from the different morphologies of neutral islands in the models.  We demonstrate this in Appendix \ref{appendix:neutralislandmorph}, by comparing the \textsc{Democratic} and \textsc{Bright-LAE-driven} models at fixed $x_{\rm HI}$. We find that the neutral islands are more clustered in the latter such that, at fixed $x_{\rm HI}$, there are significant differences in the opacity-density relation at the low-$\tau_{\rm eff}^{50}$ end. This highlights the strong degeneracy between the effects of source clustering and the global neutral fraction.  It would be difficult to constrain the source clustering from the opacity-density relation alone, without tight (better than a few percent in $x_{\rm HI}$) constraints on the global neutral fraction.  

\begin{figure*}
\begin{center}
\resizebox{0.6\linewidth}{!}{\includegraphics{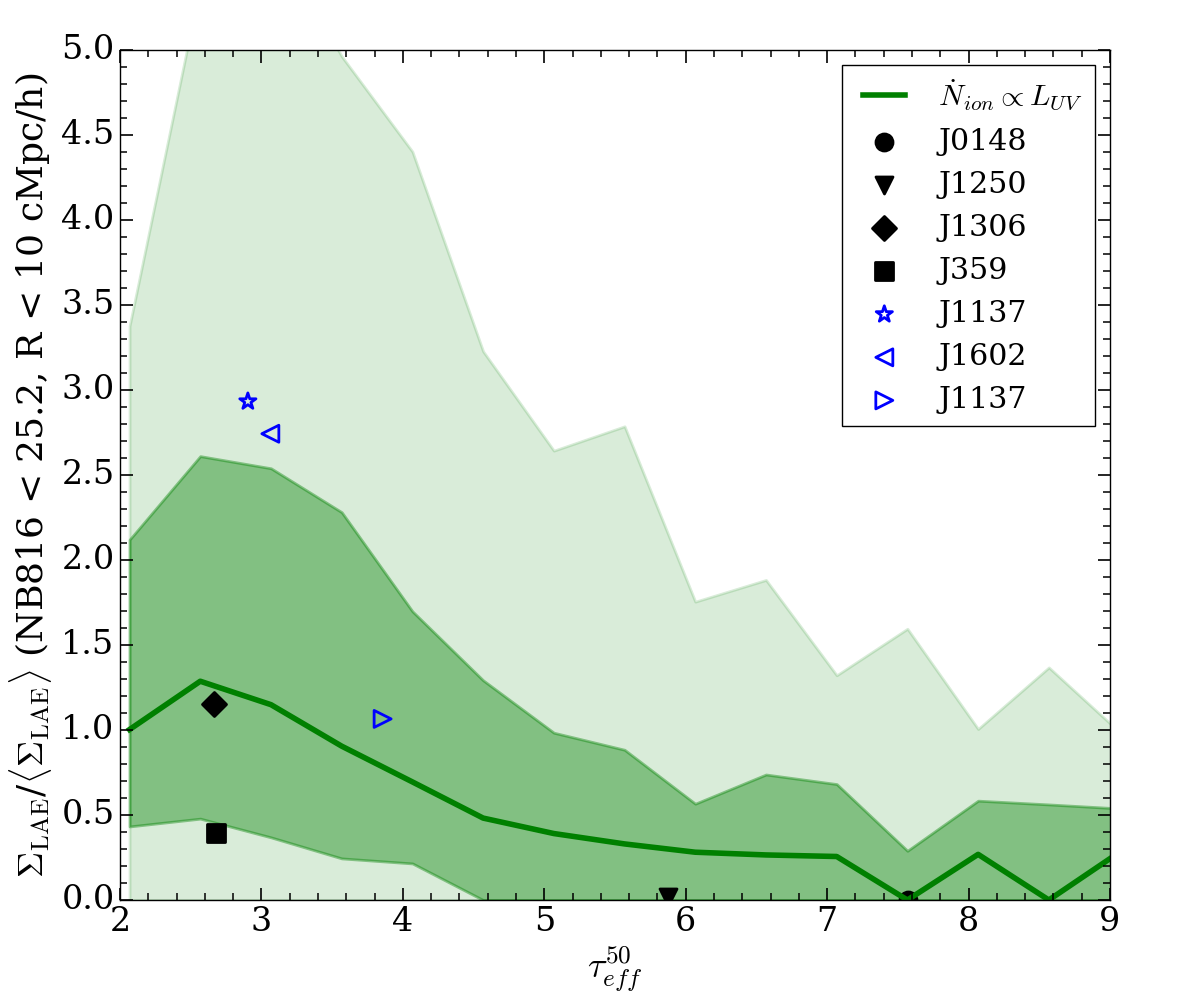}}
\end{center}
\caption{Opacity-density relation compared to an expanded set of measurements including the fields observed in Ref. \cite{Ishimoto2022}. For this comparison, we apply the limiting magnitude of NB816 $\leq 25.2$ from Ref. \cite{Ishimoto2022} for all simulated and observed fields. (All other comparisons in this paper use NB816 $\leq 25.5$ from C23.) For the fields of Ref. \cite{Ishimoto2022}, we use the $\tau^{50}_{\rm eff}$ reported in C23. For brevity we show only our our \textsc{$\Nion \propto \LUV$} model. The results are broadly similar in the other models.}
\label{fig:Ishimoto_SDvtaueff}
\end{figure*}

We return to the question of the rarity of J359 (square data point), whose central LAE surface density falls near the lower boundaries of the expected ranges from our models. C23 report 1 LAE within a circle of radius of $10~\mathrm{cMpc}/h$, although there is another LAE just outside of this circle (see their Figure 6).  To quantify how rare this sight line is, we compute the probabilities of finding $\leq 1~(2)$ LAEs within a circle of this radius in our models. We find probabilities of $\approx 4~(9)$\% in all of our models. These probabilities are likely to be sensitive to the reionization history. For instance, if reionization ended somewhat earlier than in our models, there would be fewer neutral islands in the voids and more patches of hot, recently reionized gas there.  In this case, we would expect a higher probablity of observing a field like the one toward J359.  We will demonstrate this point explicitly in \S \ref{sec:reion_timing}.

So far we have restricted our comparisons to the fields observed in C23.  Figure \ref{fig:Ishimoto_SDvtaueff} adds the three fields observed by Ref. \cite{Ishimoto2022}. For this comparison, we apply their shallower magnitude cut of NB816 $\leq 25.2$ to both our model fields and to the C23 observations. We use the $\tau^{50}_{\rm eff}$ values reported in C23 for the sight lines of Ref. \cite{Ishimoto2022}. Since the results are so similar across our source models, Figure \ref{fig:Ishimoto_SDvtaueff} shows only our \textsc{$\dot{N}_{\rm ion} \propto L_{\rm UV}$} model. The figure further demonstrates consistency between the measured sample and the model, but also highlights how deeper imaging of the fields toward transmissive sight lines in particular could help expose possible tension. For instance, the observation of more fields like that of J359 would put tension on the models.    

Lastly, we note that the opacity-density relations in Figure \ref{FIG:Sigma_vs_taueff} are markedly different from the models of Refs. \cite{Davies2016} and \cite{NasirDAloisio19}, which are shown in C23 (see their Figure 10).  In those models, the low $\tau^{50}_{\rm eff}$ sight lines are more strongly associated with over-densities.  Their median $\Sigma_{\rm LAE}(R<10~\mathrm{cMpc}/h)/\langle \Sigma_{\rm LAE} \rangle$ reach values around $2$ for $\tau_{\rm eff} \sim 2$, and the field toward J359 falls well outside of their 98\% regions.  The discrepancy likely arises from differences in the structure of the ionizing radiation backgrounds in those models. Indeed, compared to our simulations, the ionizing backgrounds from the models of Ref. \cite{NasirDAloisio19} exhibit larger fluctuations; the photoionization rate tends to be significantly higher in over-dense regions, accounting for their stronger association with low $\tau^{50}_{\rm eff}$ sight lines. In this case, the ionizing intensity fluctuations always dominate over the temperature and density fluctuations in setting the Ly$\alpha$ forest opacity. Fields like those of J359 and J1306, however, suggest that reionization temperature fluctuations may also play an important role. We will discuss this further in \S \ref{sec:GasProp}.

\subsection{Effect of Ly$\alpha$ Line Properties} \label{sec:LineProps} 

 In this section, we address the question of how the Ly$\alpha$ emission line properties affect the opacity-density relation.  For instance, could scatter in the line properties increase the scatter in the opacity-density relation, perhaps making sight lines like J359 less rare?  

\begin{figure}
\begin{center}
\resizebox{0.6\linewidth}{!}{\includegraphics{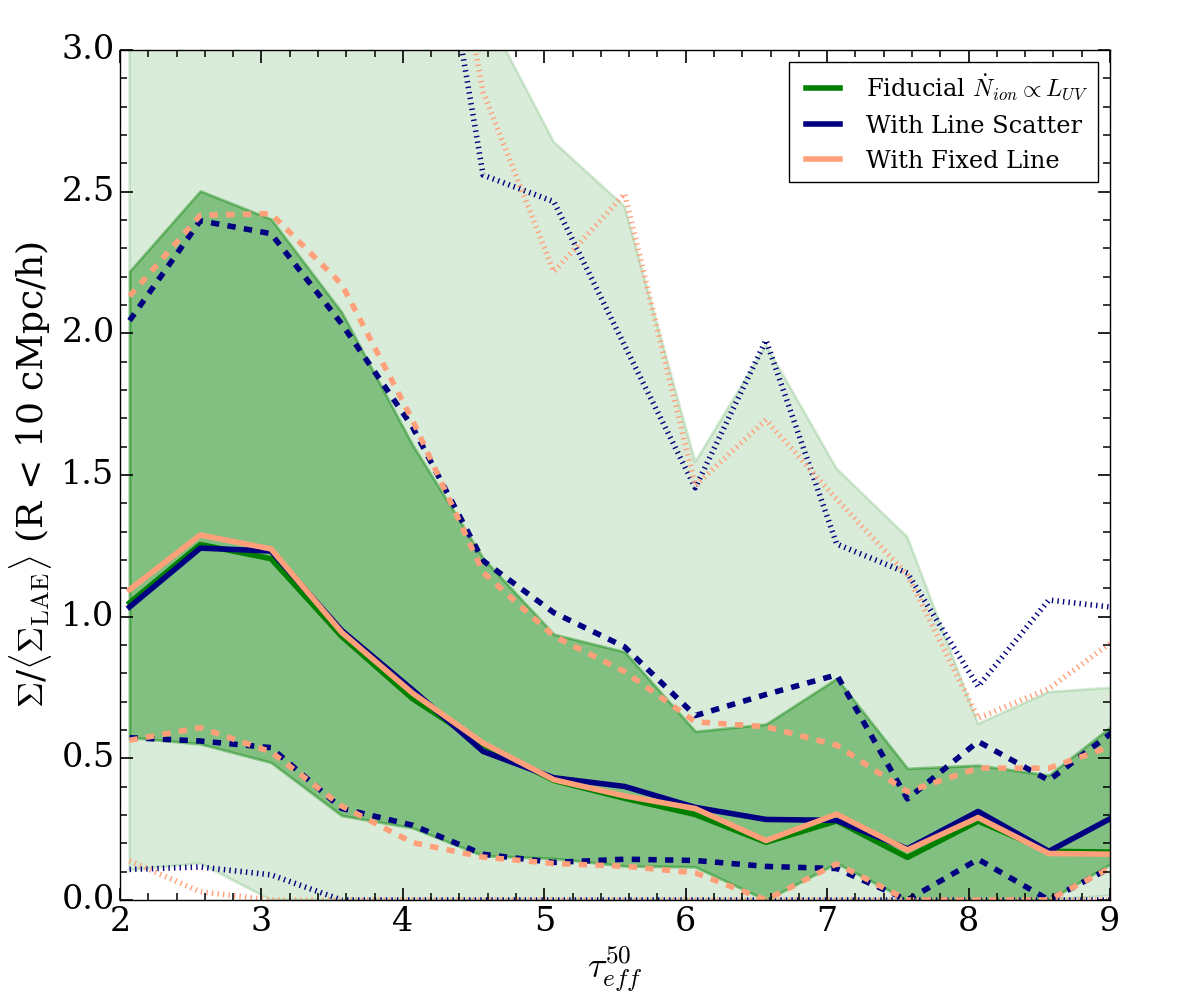}}
\end{center}
\caption{Testing how sensitive the opacity-density relation is to assumptions about the Ly$\alpha$ emission lines of narrow-band-selected LAEs. As in Figure \ref{FIG:Sigma_vs_taueff}, the green curve and shaded regions correspond to the median and 68/98\% regions for our fiducial $\dot{N}_{ion} \propto \LUV$ model.  We overlay two variations of this model.  The blue curves correspond to a variation that adds scatter to the line offset ($v_{off}$) and width ($\sigma_{v}$) according to an empirically calibrated distribution of Ref. \cite{Mason2018v1}. The yellow curves show a variation in which we hold $v_{\rm off} = 160$ km/s and $\sigma_{v} = 100$ km/s fixed.   The opacity-density relation as measured in narrow band LAE surveys is largely insensitive to the highly uncertain modelling of the $\lya$ line.}
\label{fig:lineprops}
\end{figure}

Recall that all of the models considered in the last section assume that a galaxy's rest-frame UV magnitude uniquely determines both the offset and width of its Ly$\alpha$ emission line (see \S \ref{sec:Methods}). Here we construct two test models varying this relationship. The first variation adds scatter to the relationship. For a given $M_{\rm UV}$, we draw $v_{\rm off}$ from the distribution in equation (1) of Ref. \cite{Mason2018v1}, which is motivated by a collection of observational measurements of $v_{\rm off}$ from the literature (see their Fig. 2). Since we set $\sigma_{v} = v_{\rm off}$, this procedure also implements scatter in the line width.  In the second test model, we keep $v_{\rm off}$ (and therefore also $\sigma_{v}$) constant for all galaxies to draw an extreme contrast with the first test model.   Note that the different assumptions about the line affects the visibilities of the LAEs.  LAEs for which the line offset is scattered to lower values become less likely to be visible because of IGM attenuation and inflows in the neighborhoods of galaxies.   We emphasize that the physical properties of the IGM (neutral island distribution, $\Gamma_{\rm HI}$ and $T$ fields) in these tests remain the same as before.  

Figure \ref{fig:lineprops} shows the opacity-density relations in these variations. For brevity we consider only the $\Nion \propto \LUV$ reionization model. As in Figure \ref{FIG:Sigma_vs_taueff}, the green curve and contours correspond to the median and 68/98\% regions for our fiducial $\Nion \propto \LUV$ model.  The blue and yellow solid, dashed, and dotted curves show the median and 68/98\% regions for the two model variations, as indicated in the plot legend.   The opacity-density relations are remarkably similar between the variations despite very different assumptions about the line properties. Indeed, introducing significant scatter to the line offset and width does not substantially add to the scatter in the relation, and under-densities around transmissive sight lines such as the one toward J359 do not become more likely.\footnote{We have tested this conclusion with a model implementing more extreme scatter in the line properties as well.}  In the model with line scatter we find that the probability of observing $\leq$ 1 (2) LAEs within a circle of radius $10~\mathrm{cMpc}/h$ is 3 (8) \%, compared to 4 (9) \% in the fiducial model.  We conclude that the opacity-density relation as measured from narrow band LAE surveys is broadly insensitive to the details of how the Ly$\alpha$ emission line is modeled.  On the one hand, it is reassuring that the simulation predictions are so insensitive to what we assume about the sources and the Ly$\alpha$ emission lines. On the other hand, these results again support the conclusion that it would be difficult to use the opacity-density relation alone to learn about reionization's sources.

\subsection{Cosmological environments of Ly$\alpha$ forest sight lines} \label{sec:GasProp}

\begin{table}
\begin{center}
\resizebox{0.9\columnwidth}{!}{
\begin{tabular}{c  c  c  c  c  c}
\hline 
  & $\tau^{50}_{\rm eff, min}, \tau^{50}_{\rm eff, max}$ & $\Sigma^{R<10}_{\mathrm{LAE, min}},\Sigma^{R<10}_{\mathrm{LAE, max}}$ & $N^{R<10}_{\mathrm{LAE, min}}, N^{R<10}_{\mathrm{LAE, min}}$ & N$_{\rm sample}$ \\ [0.5ex] 
 \hline
 \textsc{High-$\tau_{\rm eff}$, Low-$\Sigma$} & 7, $\infty$ & 0, 0.007 & 0, 2 & 212 \\
 \textsc{Low-$\tau_{\rm eff}$, Low-$\Sigma$} & 2.42, 2.92 & 0, 0.007 & 0, 2 & 50 \\
 \textsc{Low-$\tau_{\rm eff}$, Mean-$\Sigma$} & 2.42, 2.92 & 0.018, 0.024 & 6, 8  & 89 \\
\textsc{Low-$\tau_{\rm eff}$, High-$\Sigma$} & 2.42, 2.92 & 0.06, $\infty$ & 20, $\infty$ & 42 \\
 \hline \end{tabular}}
 \end{center}
 \vspace{-0.3cm}
 \caption{Summary of categories for Ly$\alpha$ forest sight lines used in \S \ref{sec:GasProp}. The 2nd and 3rd columns give the ranges of Ly$\alpha$ forest effective optical depths, $\tau^{50}_{\rm eff}$, and central LAE surface densities, $\Sigma_{\rm LAE}(R < 10~\mathrm{cMpc}/h)$, in units of $(\mathrm{cMpc}/$h$)^{-2}$, defining the categories. The 4th column provides the corresponding range in number of LAEs within $R < 10~\mathrm{cMpc}/h$. The last column gives the number of sight lines in the corresponding category for the \textsc{$\dot{N}_{\rm ion} \propto L_{\rm UV}$} model.}
 \label{tab:op_dens_groups}
\end{table}

We turn our attention to the physical conditions that give rise to the low and high $\tau_{\rm eff}^{50}$ sight lines in our simulations. To this end, we find it useful to define four categories detailed in Table \ref{tab:op_dens_groups}. The field toward J0148 fits into the \textsc{High-$\tau_{\rm eff}$, Low-$\Sigma$} category, while J359 fits into the \textsc{Low-$\tau_{\rm eff}$, Low-$\Sigma$} category. We also introduce the \textsc{Low-$\tau_{\rm eff}$, Mean-$\Sigma$} and \textsc{Low-$\tau_{\rm eff}$, High-$\Sigma$} categories to span the variation in low-$\tau^{50}_{\rm eff}$ sight lines. The 2nd and 3rd columns give the ranges of $\tau^{50}_{\rm eff}$ and $\Sigma_{\rm LAE}(R < 10~\mathrm{cMpc}/h)$ used to define the categories.  For reference, the 4th column gives the corresponding number of LAEs within $R < 10~\mathrm{cMpc}/h$, and the last column gives the number of fields in each category. In this section we show results from the \textsc{$\dot{N}_{\rm ion} \propto L_{\rm UV}$} model, but the qualitative discussion applies to all of the models.     

\begin{figure*}
\begin{center}
\resizebox{\linewidth}{!}{\includegraphics{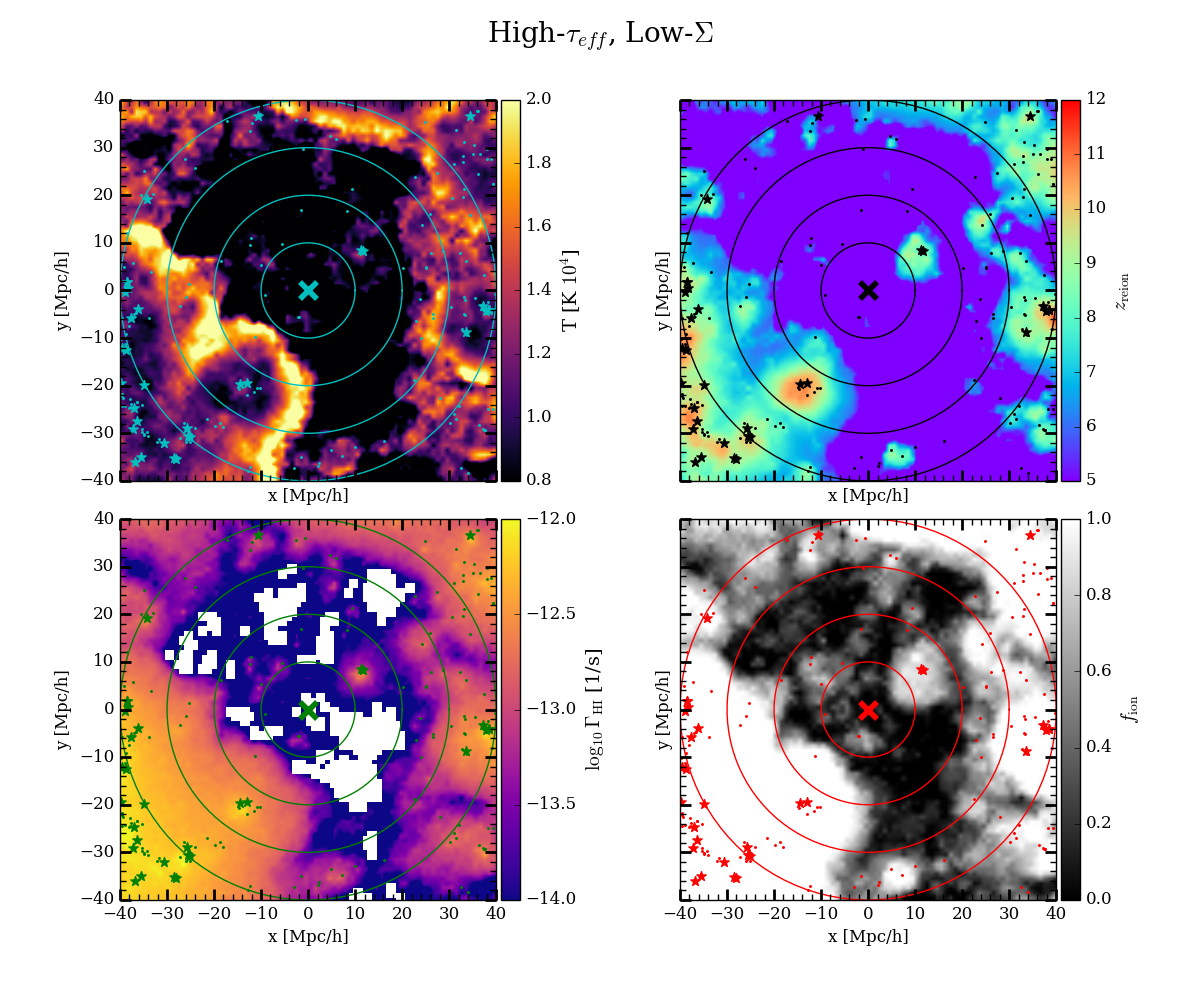}}
\end{center}
\vspace{-1cm}
\caption{Physical environment of a \textsc{High-$\tau_{\rm eff}$, Low-$\Sigma$} field, such as the one toward J0148. All plots in this section use the \textsc{$\dot{N}_{\rm ion} \propto L_{\rm UV}$} model, but the qualitative discussion applies to all of the models.  Starting in the upper left and moving clockwise, the maps show temperature ($T$), redshift of reionization ($z_{\mathrm{reion}}$), ionized fraction ($x_{\mathrm{ion}}$), and hydrogen photoionization rate ($\Gamma_{\rm HI}$). Distances on the axes are in cMpc$/h$ and slices have been averaged over a depth of 12 cMpc/$h$.  The small dots show galaxies (ionizing sources), while the stars show ``observed" LAEs in our mock surveys, i.e. those passing the magnitude and color cuts.  The large ``X'' denotes the centers of the fields. Concentric circles correspond to radial increments of 10 cMpc$/h$.  \textsc{High-$\tau_{\rm eff}$, Low-$\Sigma$} fields are characterized by large neutral structures near the center of the field and low \GammaHI\ values. }
\label{fig:HOLD}
\end{figure*}

 Let us begin with Figure \ref{fig:HOLD}, which visualizes a \textsc{High-$\tau_{\rm eff}$, Low-$\Sigma$} field. Starting at the top left and moving clockwise, we show the gas temperature, redshift of reionization ($z_{\rm reion}$), ionized fraction ($x_{\rm ion}$), and hydrogen photoionization rate (\GammaHI). The slices are $80\times80$ (cMpc/$h$)$^2$ and are averaged over a depth of 12 cMpc/$h$.  All maps are taken at $z=5.7$. The concentric circles represent radial increments of 10 cMpc/$h$.  In each panel, the smaller dots show galaxies/ionizing sources while the larger stars show ``observed" LAEs in the field, i.e. those that make the magnitude and color cuts described in \S \ref{sec:Methods}. The large ``X'' denotes the centers of the fields.  This field is characterized by a dearth of galaxies and LAEs within a radius of $\sim 20$ cMpc/$h$, and neutral island structure covering much of the inner field, depicted in black in the $x_{\rm ion}$ map, as well as in white in the \GammaHI\ map. Indeed, in general we find that \textsc{High-$\tau_{\rm eff}$, Low-$\Sigma$} sight lines are almost always centered near large neutral structures.  Note that much of the \GammaHI\ field adjacent to the neutral islands reaches values as low as $\sim 10^{-14}~\mathrm{s}^{-1}$ owing to the lack of sources and the effects of shadowing by the islands (see Ref. \cite{NasirDAloisio19} for more discussion on this shadowing effect). The gas at the boundaries of the neutral islands have $T\sim 20,000$ K, consistent with the hot temperatures behind ionization fronts \cite{DAloisio2019}.

\begin{figure*}
\begin{center}
\resizebox{\linewidth}{!}{\includegraphics{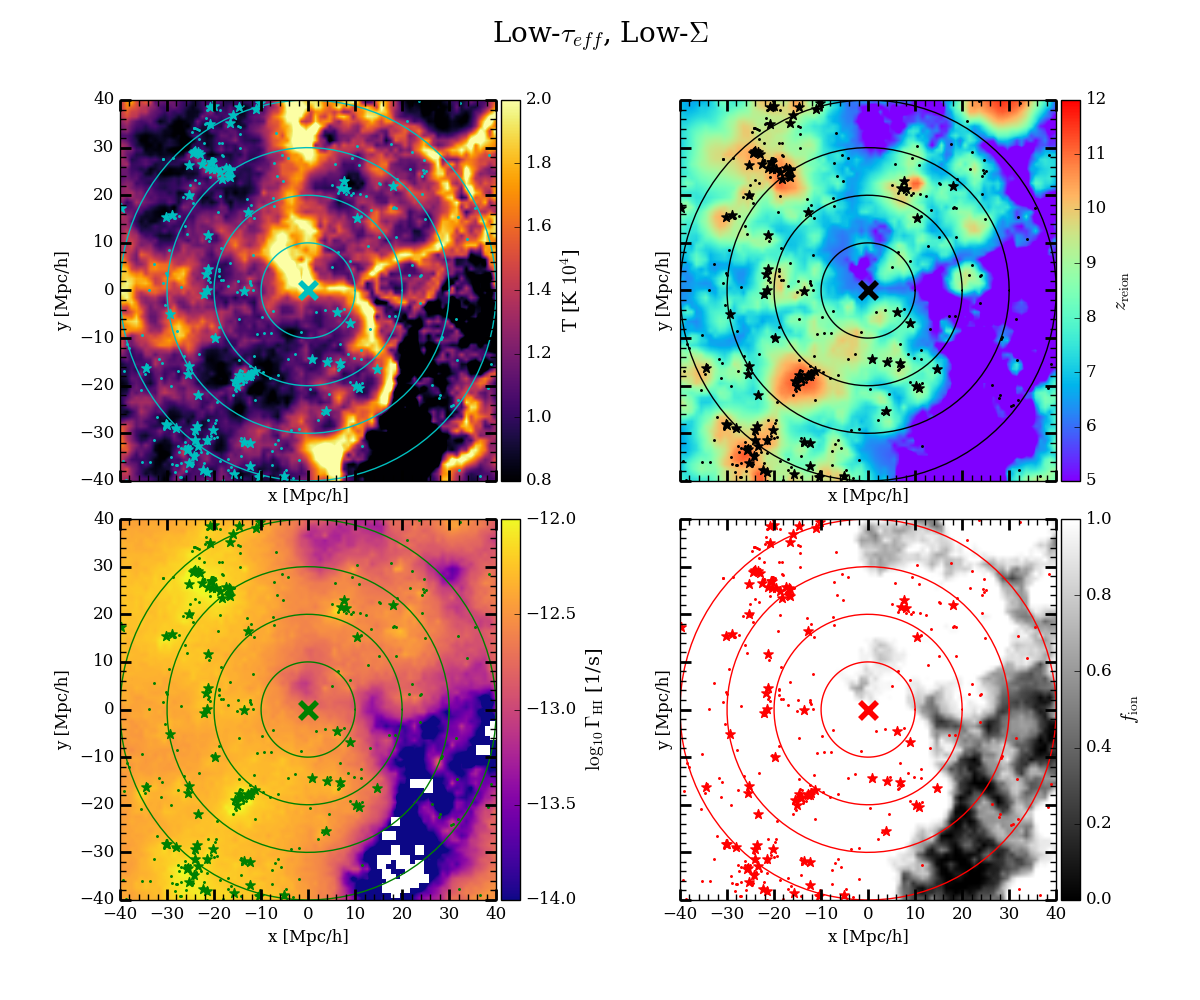}}
\end{center}
\vspace{-1cm}
\caption{Same as in Figure \ref{fig:HOLD}, but for a \textsc{Low-$\tau_{\rm eff}$, Low-$\Sigma$} field. The field around J359 fits into this category. These fields are characterized by hot ($T\sim 20,000$ K), recently reionized, under-dense gas in their centers.  Another important characteristic is that there are clear lines of sight from the center to over-densities of sources, such that $\Gamma_{\rm HI}$ is not suppressed there.}
\label{fig:LOLD}
\end{figure*}

Next we examine the \textsc{Low-$\tau_{\rm eff}$, Low-$\Sigma$} field shown in Figure \ref{fig:LOLD}. This field contains more LAEs than the one in Figure \ref{fig:HOLD}, but the center of the field is also under-dense in sources/LAEs. Together, the maps reveal what gives rise to the high Ly$\alpha$ forest transmission at the center of the map.  The center intersects a cosmological under-density that was recently reionized, as indicated by the adjacent neutral island and the hot $\sim20,000$K temperatures of the gas. In fact, as we will demonstrate below, most sight lines in the \textsc{Low-$\tau_{\rm eff}$, Low-$\Sigma$} category exhibit enhanced central temperatures from recently reionized gas.  Another important characteristic is that there are clear lines of sight from the center to over-densities of sources (in this case to the upper and lower left), such that \GammaHI\ is not suppressed by shadowing effects.  The combination of these factors leads to enhanced forest transmission at the center of the field.  

\begin{figure*}
\begin{center}
\resizebox{\linewidth}{!}{\includegraphics{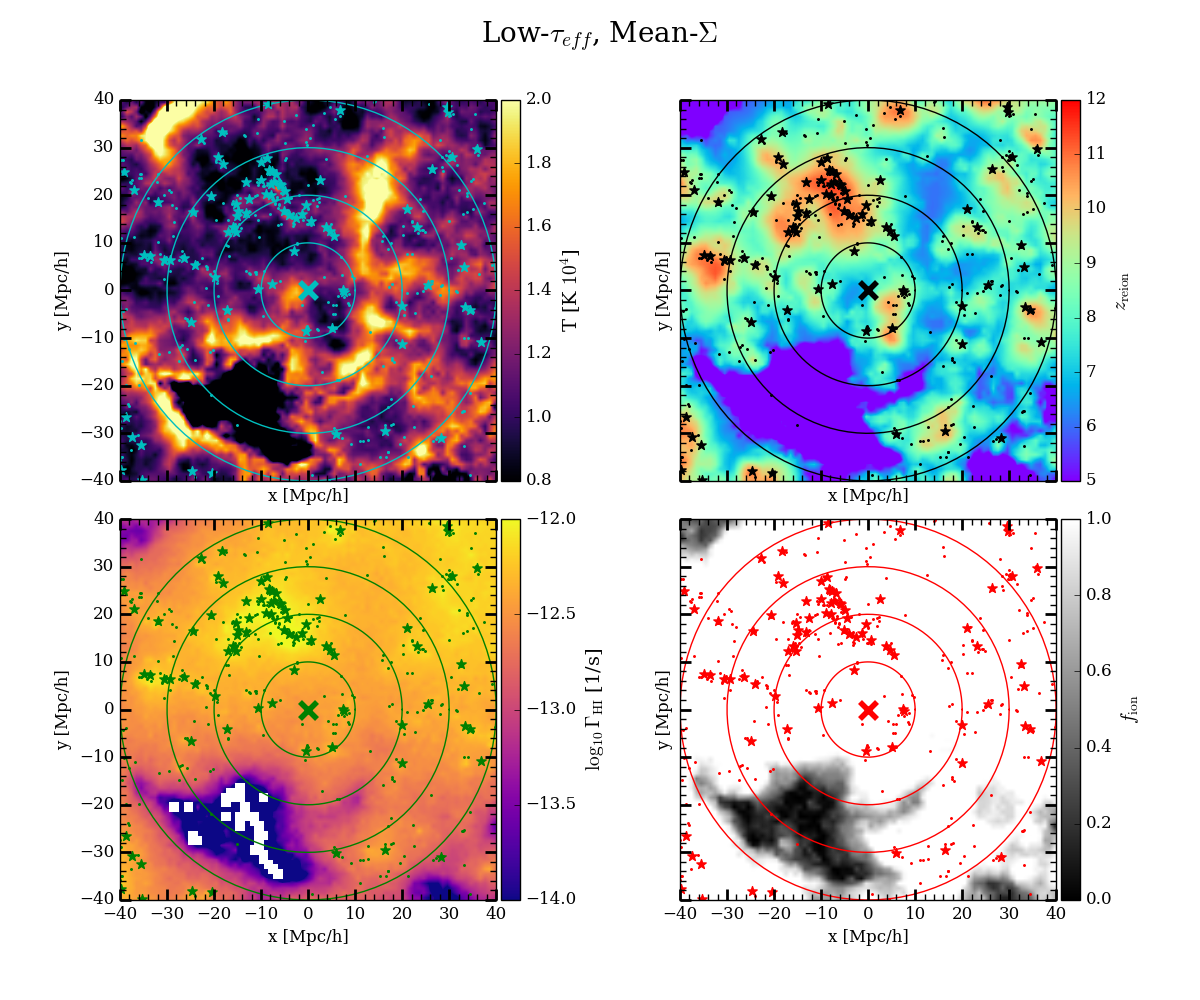}}
\end{center}
\vspace{-1cm}
\caption{Same as in Figure \ref{fig:HOLD}, but for a \textsc{Low-$\tau_{\rm eff}$, Mean-$\Sigma$} field. The fields in this category are more mixed in the effects of $\Gamma_{\rm HI}$ versus $T$.  In some fields, the proximity to an over-density of sources (and hence higher values of $\Gamma_{\rm HI}$) is the larger effect for enhancing the forest transmission. In other cases, the hotter temperatures from recently reionized gas is the larger effect.  In this figure we show an example of the former.}  
\label{fig:LOMD}
\end{figure*}

The majority of fields around low-$\tau^{50}_{\rm eff}$ sight lines have $\Sigma_{\rm LAE}(R < 10~\mathrm{cMpc}/h)$ close to the mean value (see Fig. \ref{FIG:Sigma_vs_taueff}). Figure \ref{fig:LOMD} visualizes an example of a \textsc{Low-$\tau_{\rm eff}$, Mean-$\Sigma$} field.  The most distinct feature is the large over-density of sources $20$ cMpc/$h$ northwest of the center. Compared to the \textsc{Low-$\tau_{\rm eff}$, Low-$\Sigma$} field in Figure \ref{fig:LOLD}, $\Gamma_{\rm HI}$ plays a larger role in enhancing the Ly$\alpha$ forest transmission.  The central region is situated between the source over-density and the reionizing neutral island in the southwest quadrant. Examining other fields in our \textsc{Low-$\tau_{\rm eff}$, Mean-$\Sigma$} category, we find some like the one in Figure \ref{fig:LOMD}, but others in which the hotter temperatures from recently reionized gas has a more prominent effect. The diversity of this category highlights that both $\Gamma_{\rm HI}$ and $T$ can play important roles.  

\begin{figure*}
\begin{center}
\resizebox{\linewidth}{!}{\includegraphics{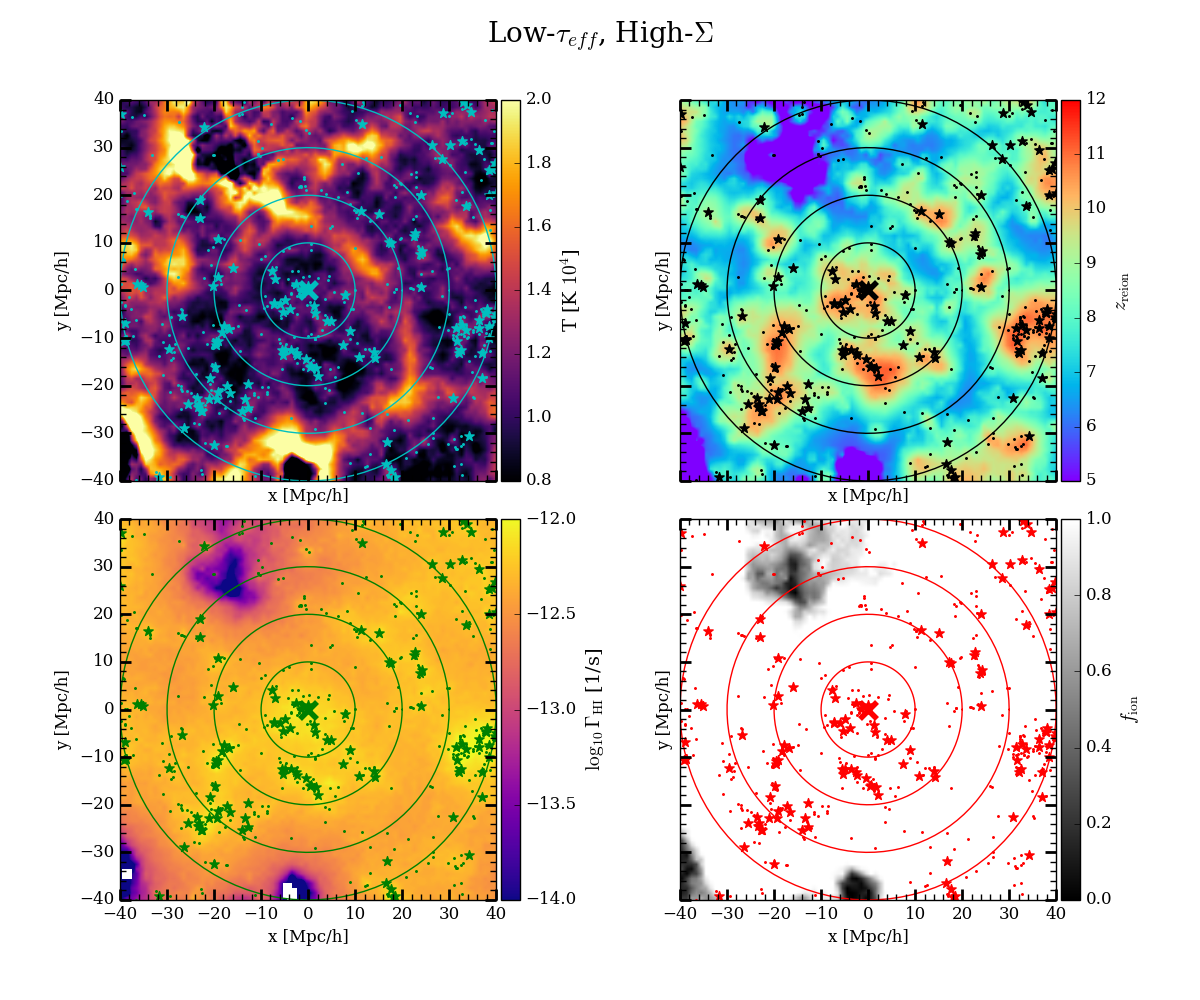}}
\end{center}
\vspace{-1cm}
\caption{Same as in Figure \ref{fig:HOLD}, but for a \textsc{Low-$\tau_{\rm eff}$, High-$\Sigma$} field.  The intense ionizing radiation in an over-density of sources is what drives the forest transmission in these fields. Such over-densities were reionized earlier, and therefor exhibit cooler gas temperatures.}
\label{fig:LOHD}
\end{figure*}

Lastly, we consider a \textsc{Low-$\tau_{\rm eff}$, High-$\Sigma$} field in Figure \ref{fig:LOHD}.  The center of the field exhibits a strong over-density of sources and correspondingly enhanced $\Gamma_{\rm HI} \sim 10^{-12}$ s$^{-1}$.  The $z_{\rm reion}$ field indicates that the center was one of the first regions to be reionized in the simulation at around $z=10-12$. As such, this region has cooled to temperatures of $T\sim 8,000$ K by $z=5.7$.  However, the high $\Gamma_{\rm HI}$ values overcome the cosmological over-density and colder temperatures to produce enhanced Ly$\alpha$ forest transmission at the center of the map. We note that these types of fields dominate the transmissive end of the opacity-density relation in the strong intensity fluctuation models of Refs. \cite{Davies2016,NasirDAloisio19}.

\begin{figure}
\begin{center}
\resizebox{0.9\columnwidth}{!}{\includegraphics{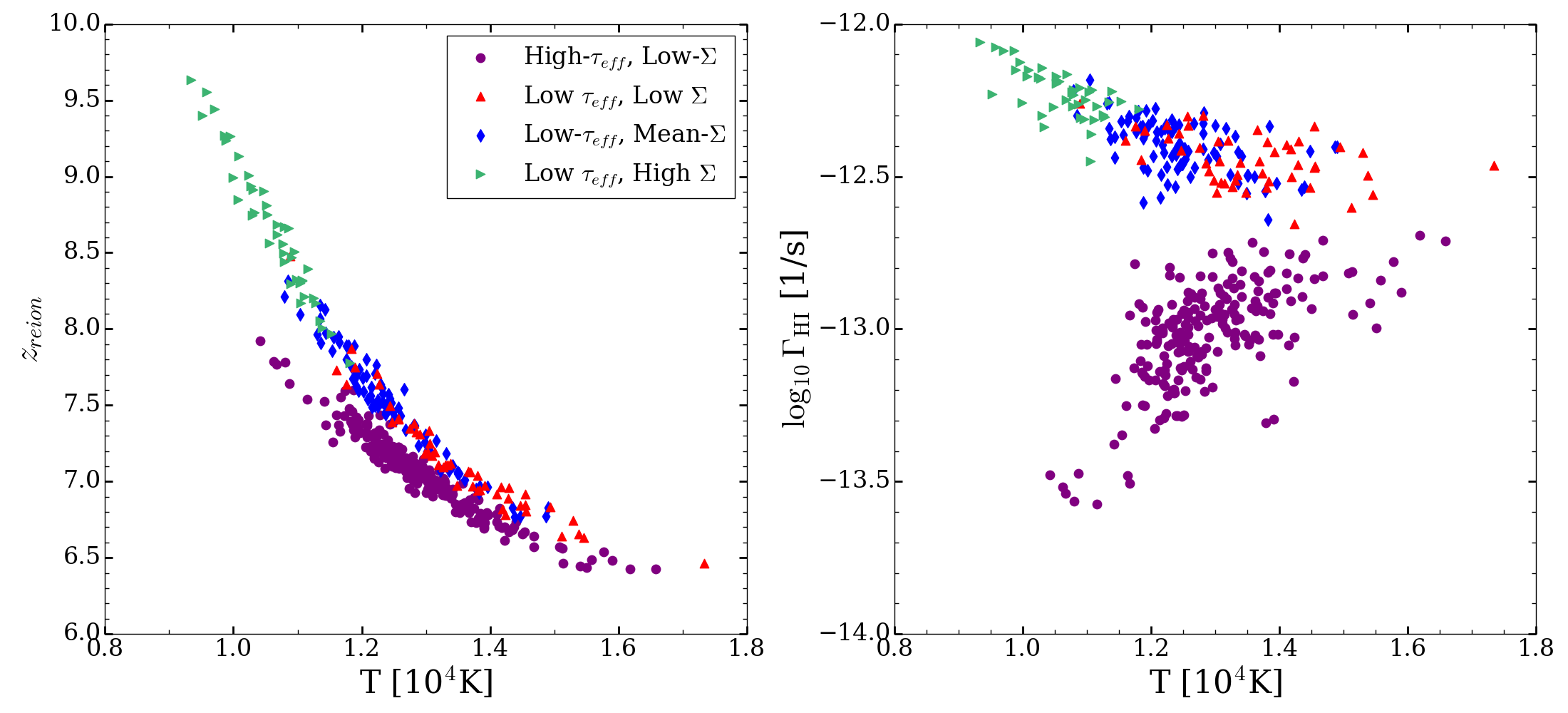}}
\end{center}
\caption{Physical properties at the centers of fields in our four categories (defined in Table \ref{tab:op_dens_groups}). We average the temperature ($T$), redshift of reionization ($z_{\rm reion}$), and hydrogen photoionization rate ($\Gamma_{\rm HI}$) in fully ionized cells within a cylinder of radius 10 cMpc$/h$ and depth of 30cMpc$/h$, centered on the simulated Ly$\alpha$ forest sight line. The depth roughly corresponds to the narrow band filter NB816 width. Each data point corresponds to a single field.  The left panel displays the $z_{\mathrm{reion}}$-$T$ plane and the right panel displays the $\Gamma_{\rm HI}$-$T$ plane. Fields from the different categories occupy different regions of these planes, depending on the physical conditions giving rise to the Ly$\alpha$ forest transmission or absorption. }
\label{fig:temp_vs_zreion}
\end{figure}

Figure \ref{fig:temp_vs_zreion} attempts to show more quantitatively the physical conditions giving rise to the fields in our four categories. Here we consider the central values of $T$, $z_{\rm reion}$, and $\Gamma_{\rm HI}$.  We average these quantities in fully ionized cells within a cylinder of radius 10 cMpc/$h$ and depth of 30cMpc/$h$, centered on the simulated Ly$\alpha$ forest sight line. The depth was chosen to span roughly the full width at half maximum (FWHM) of the narrow band filter, NB816. The data points in Figure \ref{fig:temp_vs_zreion} show each field within the four categories in the $z_{\rm reion}-T$ and $\Gamma_{\rm HI}-T$ planes. 

Consider first the \textsc{High-$\tau_{\rm eff}$, Low-$\Sigma$} fields shown as the purple circles in Figure \ref{fig:temp_vs_zreion}.  As expected from our previous discussion, their central regions contain gas that was reionized more recently, with average values $z_{\rm reion}\lesssim 8$. Naturally the average central $T$ is on the hotter side of the range. (Recall that our averaging procedure excludes the neutral islands.) On the other hand, the right panel shows that the average $\Gamma_{\rm HI}$ in this gas is lower than in the other categories.  These characteristics are consistent with our previous discussion that the \textsc{High-$\tau_{\rm eff}$, Low-$\Sigma$} fields contain neutral islands. The gas bordering the neutral islands are recently reionized and hot, but the $\Gamma_{\rm HI}$ is suppressed in regions between the neutral islands because of shadowing effects.  Next, consider the \textsc{Low-$\tau_{\rm eff}$, Low-$\Sigma$} fields (red triangles).  These are also among the hot, more recently reionized regions. Note, however, that the average central $\Gamma_{\rm HI}$ values are on the higher side of the range. Again, this is consistent with our previous discussion on how the low densities, hot temperatures, and higher values of $\Gamma_{\rm HI}$ combine to produce enhanced Ly$\alpha$ forest transmission.  Similar remarks apply to the \textsc{Low-$\tau_{\rm eff}$, Mean-$\Sigma$} fields (blue diamonds), although $\Gamma_{\rm HI}$ is likely to play a larger role in producing the forest transmission for these sight lines.  Lastly, the \textsc{Low-$\tau_{\rm eff}$, High-$\Sigma$} fields (green triangles) clearly correspond to the colder regions with enhanced $\Gamma_{\rm HI}$ that were reionized earliest.  The ionizing background in those regions is strong enough to overcome the local over-density and colder temperatures to produce enhanced transmission. These results shed light into the physical conditions giving rise to the fields in our samples.

\section{Sensitivity to the Reionization History} 
\label{sec:reion_timing}

\begin{figure*}
\begin{center}
\resizebox{0.8\columnwidth}{!}{\includegraphics{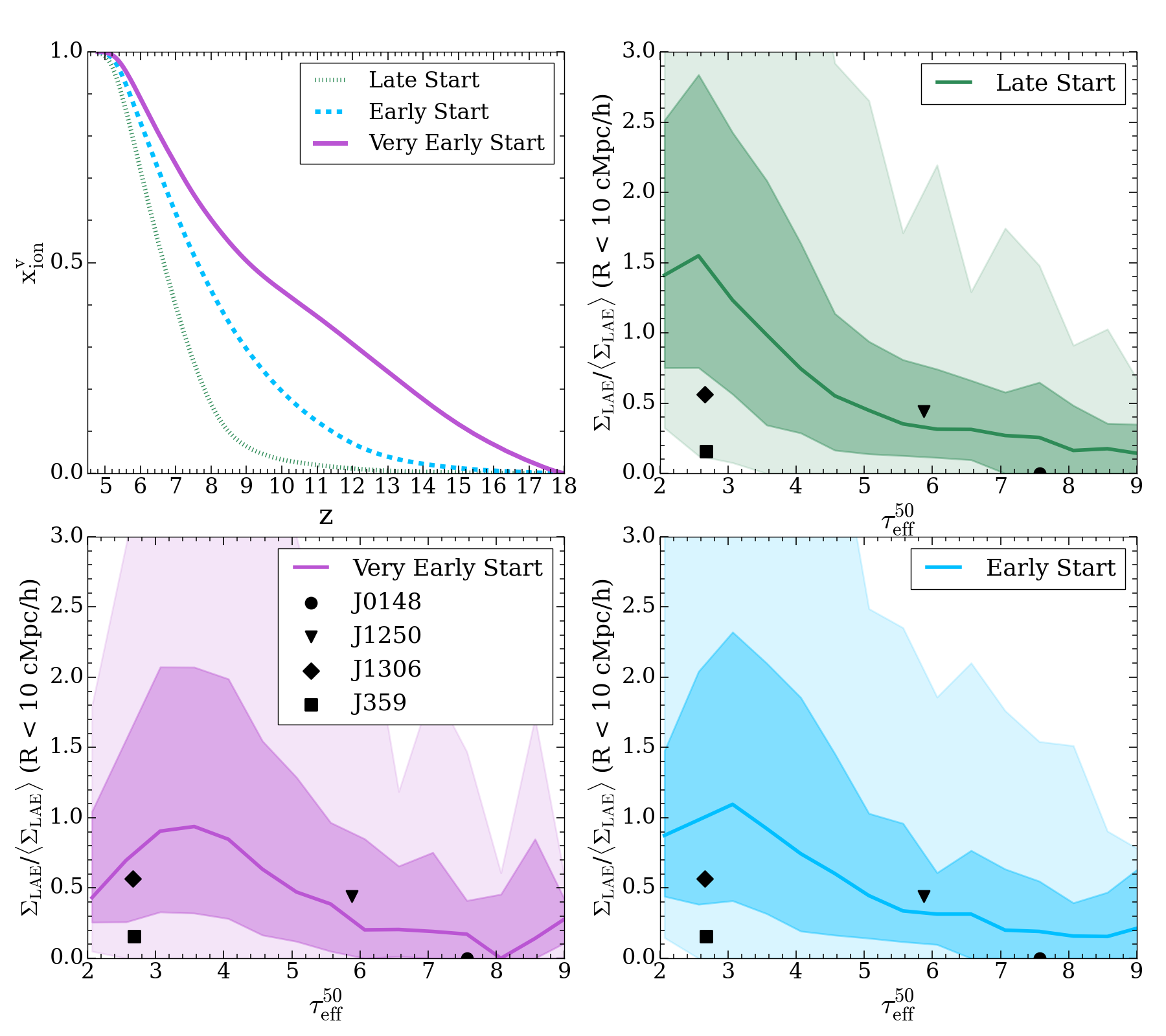}}
\end{center}
\caption{Effect of the reionization history on the opacity-density relation. The top-left panel shows the reionization histories for the three additional simulations considered in \S \ref{sec:reion_timing}, as annotated in the legend.  All three models have been calibrated to match measurements of the mean Ly$\alpha$ forest transmission at $z\leq 6$, but they have very different reionization histories overall.  At $z=5.7$, the \textsc{Late Start}, \textsc{Early Start}, and \textsc{Very Early Start} models have global neutral fractions of $x_{\rm HI} = 17$, $10$, and $6$ \%, respectively. When the neutral fraction is lower, the larger physical cross section of hot, recently reionized gas in the voids increases the likelihood that the most transmissive forest sight lines intersect galaxy under-densities. Hence the low-$\tau^{50}_{\rm eff}$ end of the opacity-density relation is particularly sensitive to the timing of reionization's ending phases.}
\label{fig:reion_timing}
\end{figure*}

In \S \ref{sec:opac_dens}, we alluded to the idea that the opacity-density relation could be sensitive to the reionization history.  We demonstrate this here.  For this task we use three additional reionization simulations with varied reionization histories. To save computational time, the simulations used in this section were run with monochromatic RT. Photons were assigned an energy of $E_\gamma = 19$ eV to match the frequency-averaged H photoionization cross section of a power-law spectrum with $J_\nu \propto \nu^{-1.5}$ between 1 and 4 Ry. We adopt a source model similar to the \textsc{$\dot{N}_{\rm ion} \propto L_{\rm UV}$} used in previous sections.\footnote{There is one significant difference in the source modeling. Here we will explore models in which reionization begins significantly earlier than in our fidicial simulations. To maximize the number of halos in the box at these high redshifts ($z>15$), we use halos down to a minimum halo mass of $M_{\rm min} = 1\times 10^9~h^{-1}$ M$_{\rm \odot}$ in this section. Our halo mass function is only 50\% complete at this lower minimum mass.  We note, however, that in reality the drop-off in star formation efficiency cannot be perfectly sharp in halo mass anyways.}    The top-left panel of Figure \ref{fig:reion_timing} shows the reionization histories of the three simulations.  As before, the simulations have been calibrated to agree with the observed mean Ly$\alpha$ forest transmission at $z<6$. (We do not show a plot here for brevity.) Note that the reionization histories differ significantly between the models, particularly in the earlier phases at $z>7$.  From here on we adopt the following names for the models shown with the green/dotted, cyan/dashed, and purple/solid curves: \textsc{Late Start}, \textsc{Early Start}, and \textsc{Very Early Start}, respectively.  Importantly for this discussion, the variations in the $z>7$ reionization histories result in different neutral fractions at $z=5.7$ at roughly fixed Ly$\alpha$ forest transmission.     The \textsc{Late Start}, \textsc{Early Start}, and \textsc{Very Early Start} models have global neutral fractions at $z=5.7$ of $x_{\rm HI} = 17$, $10$, and $6$ \%, respectively. We emphasize that these models have a fixed source clustering since they all adopt the \textsc{$\dot{N}_{\rm ion} \propto L_{\rm UV}$} model.

The remaining panels in Figure \ref{fig:reion_timing} compare the opacity-density relations between the three models. As anticipated, there are significant differences in the opacity-density relations at the low-$\tau_{\rm eff}^{50}$ end.  Under-densities around highly transmissive sight lines at $z=5.7$ (such as in J1306 and J359) become significantly more likely if the global neutral fraction is lower, in better agreement with the data.  For instance, in the \textsc{Late Start} model, we find a probability of just 3\% for observing $\leq 1$ LAE within a circle of radius $R = 10~\mathrm{cMpc}/h$ (such as in J359).  This probability increases to 9\% in the \textsc{Early Start} model, and to 15\% in the \textsc{Very Early Start} model. The increase in probability occurs because the likelihood of intersecting hot, recently reionized gas in the voids increases if the neutral islands are smaller, i.e. when the global neutral fraction is lower.  It follows from our discussion in \S \ref{sec:GasProp} that the likelihood of observing sight lines like J359 increases under these conditions.

Aside from the poor statistics of the current opacity-density measurements, some significant caveats prevent us from drawing the conclusion that the \textsc{Very Early Start} model is favored. There are other things besides the early ionization in this model that could also result in a lower neutral fraction at $z=5.7$ for a fixed Ly$\alpha$ forest flux. For instance, one effect likely at play is the numerical convergence of the Ly$\alpha$ forest in our simulations. As mentioned in \S \ref{sec:Methods}, we have applied the convergence correction factors in Table A1 of Ref. \cite{DAloisio2018} to our simulated Ly$\alpha$ forests. As shown in Ref. \cite{Cain2023}, for a fixed Ly$\alpha$ forest flux, this procedure has the effect of requiring a larger neutral fraction; reionization has to end later than a simulation without the correction procedure. While it is clear that some correction for resolution is needed, it is possible that we have over-corrected our simulations. Table A1 of Ref. \cite{DAloisio2018} suggests that, if anything, our correction factors may not be large enough.  But those factors were obtained from a suite of convergence tests in smaller boxes with $L=25~\mathrm{cMpc}/h$, and it is unclear how the correction factors would change in a larger box.  Other effects that could lower the $z=5.7$ neutral fraction at fixed Ly$\alpha$ forest flux include the spectra and clustering of the ionizing sources. Ref. \cite{Cain2023} found that, for a fixed Ly$\alpha$ forest flux evolution, reionization can end by as much as $\Delta z_{\rm end} \sim 0.5$ earlier in models with a harder ionizing spectrum and less biased sources.  It is possible that some combination of these effects would remove the requirement of significant ionization at $z>10$ that our models suggest. This discussion highlights some of the key uncertainties and degeneracies that would need to be addressed before one could draw robust inferences on reionization.

In summary, we have shown that the frequency of sight lines like J359 is sensitive to the global neutral fraction at $z=5.7$. Indeed, if more under-dense fields are observed around the most transmissive Ly$\alpha$ forest sight lines, the frequency of these sight lines could be used to put model-dependent upper limits on the neutral fraction.

\section{Conclusion} \label{sec:conclusion}

Observations of quasar absorption spectra provide strong evidence that the ending phases of reionization extended below $z=6$.  The opacity-density relation is a key observational test of this scenario. Using narrow-band surveys of $z\approx 5.7$ LAEs centered on quasar sight lines, C23 found that two of the most transmissive forest segments at this redshift intersect under-densities in the galaxy distribution.  For example, the field of quasar J359 has just one LAE within a radius of $10$ cMpc/$h$. These findings appear to be tension with the late reionization models of Ref. \cite{NasirDAloisio19}, which predict that the vast majority of the most transmissive forest segments should intersect galaxy over-densities, where the ionizing background is strongest.    In this paper, we have used RT simulations to explore in more detail the opacity-density relation predicted by late reionization models.

We found that the opacity-density relation's transmissive end is sensitive to the amount of neutral gas in the voids, as well as its morphology, where the latter is set by the clustering of reionization sources. These effects are, however, largely degenerate with each other.
We demonstrated this with four models that make very different assumptions about reionization's sources.  For instance, in one model, the {\it observed} LAEs are the sole leakers of ionizing radiation. In another model, the sources are predominantly the faintest galaxies in the UV luminosity function, well below current detection limits. We found that the four models yield very similar opacity-density relations when their reionization histories have been calibrated to match Ly$\alpha$ forest mean flux measurements at $z < 6$.  At fixed mean flux, the models have different neutral fractions, which largely offsets the differences in neutral island morphologies caused by source clustering.  

We examined the ionization and thermal states of intergalactic gas in the fields of the most transmissive and opaque forest segments in our simulations. As anticipated, the fields of highly opaque segments with $\tau_{\rm eff } \geq 7$ are likely to be centered on a cluster of neutral islands.   Highly transmissive forest segments that intersect galaxy under-densities (like J359) are typically associated with regions that were recently reionized. In this case, the Ly$\alpha$ forest transmission is boosted by the low cosmological density and hot temperature of the recently reionized gas.   On the other hand, highly transmissive forest segments that intersect galaxy over-densities are typically associated with the earliest regions to re-ionize.  There, the forest transmission owes to the strong ionizing radiation background from the over-density of sources. Lastly, highly transmissive sight lines that intersect average galaxy densities (the majority of transmissive sight lines in our simulations) originate from a range of environments.  Sometimes the hot temperatures of recently reionized gas boost the transmission; other times, it is the enhanced $\Gamma_{\rm HI}$ from proximity to an over-density of ionizing sources.   

Highly transmissive $z\approx 5.7$ forest segments that intersect galaxy under-densities are more common in our RT simulations than in the models of Ref. \cite{NasirDAloisio19}. To quantify the prevalence of such fields, we calculated the probability of observing a sight line like that of J359.  We found that the probability of observing $\leq 1~(2)$ LAEs within a radius of $10$ cMpc/$h$ from a forest segment with $\tau^{50}_{\rm eff} \approx 2.67$ is $\approx 4~(9) \%$ in the four source models mentioned above. (Again, that the probabilities are the same for these models owes to the degeneracy between neutral fraction and neutral island morphology, and the calibration to the observed mean forest flux.) We conclude that fields like the one toward J359 are rare in our late reionization simulations, but nonetheless compatible with them.  Interestingly, we found that the frequency of such fields is sensitive to the exact timing of reionization for fixed source clustering. For instance, we constructed two models with different neutral fractions at $z=5.7$, but adopting the same source clustering model. Both models have nearly identical Ly$\alpha$ forest mean flux evolutions at $z<6$, in agreement with measurements.  The probability of observing a field like J359 is $15\%$ in one model with neutral fraction $x_{\rm HI} = 5\%$ at $z=5.7$, three times more likely than in the other model with $x_{\rm HI} =  15\%$.  The frequency of fields like J359 could in principle be a useful complementary probe of the global neutral fraction in the tail of reionization. 
Lastly, we tested whether the opacity-density relation measured in narrow-band LAE surveys depends on assumptions about the Ly$\alpha$ emission lines of the LAEs (e.g. scatter in the line offsets and widths).  We found that the relation is insensitive to wide variations in the line properties.  This demonstrates that the predicted narrow-band opacity-density relations from simulations are robust to some of the most uncertain modeling choices. 

In this work we have focused exclusively on the type of narrow-band LAE survey carried out in Refs. \cite{Becker2018, Christenson2021, Christenson2023}. The \textsc{EIGER} \cite{EIGER2023} and \textsc{ASPIRE} \cite{2023ApJ...951L...4W} programs are conducting spectroscopic surveys with JWST to map the large-scale structure around a combined total of 31 high-$z$ quasar sight lines.  This will provide an unprecedented view into the relationship between the galaxy population and IGM transmission near the epoch of reionization.  Future work should address the question of what information about reionization and its sources can be extracted from these rich data sets.

\acknowledgments
The authors thank Hy Trac for providing the hydrodynamics and N-body codes that were used to run the cosmological simulations in this paper.  A.D.'s group was supported by grants NASA 19-ATP19-0191, NSF AST-2045600, and
JWSTAR-02608.001-A. All computations were made possible by NSF ACCESS allocations TG-PHY210041, TG-PHY230063, and TG-PHY240011, and by the NASA HEC Program through the NAS Division at Ames Research Center

\appendix
\section{Insensitivity to Ray Tracing Noise}
\label{app:ray_noise}

\begin{figure}
\resizebox{\linewidth}{!}{\includegraphics{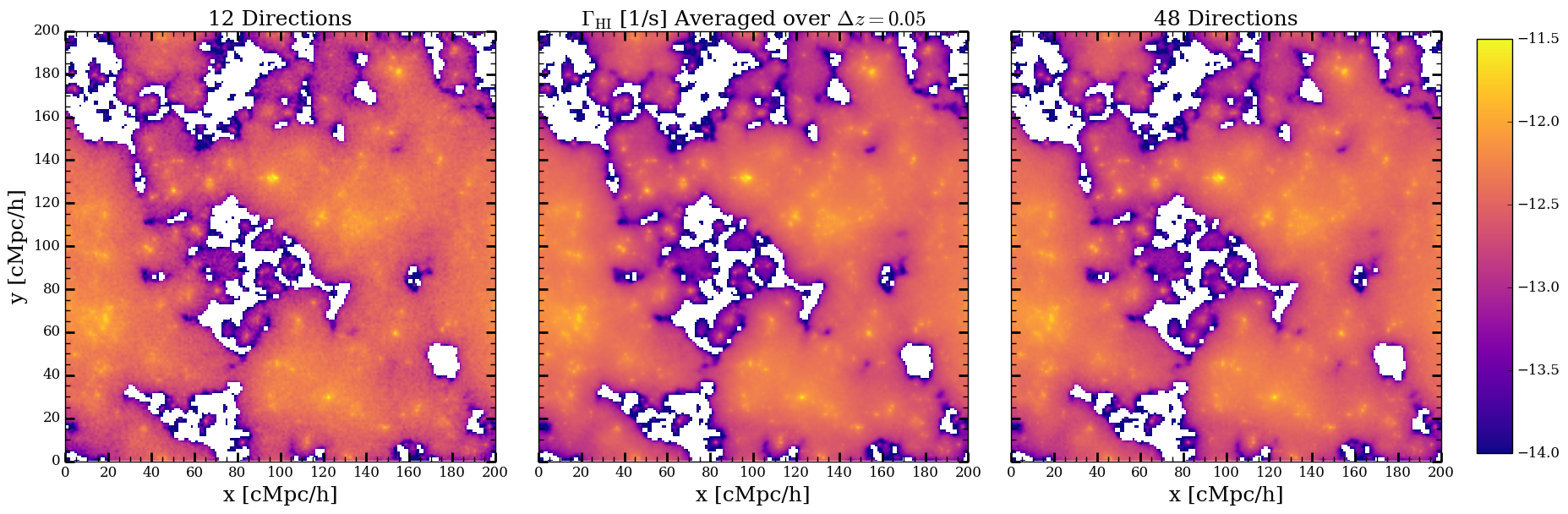}}
\caption{Ray tracing noise in RT simulations of reionization. {\it Left :} cross section through the $z=5.7$ $\Gamma_{\rm HI}$ field in our \textsc{$\dot{N}_{\rm ion} \propto L_{\rm UV}$} model. The cross section is one RT cell thick ($\Delta x = 1~\mathrm{Mpc}/h$), and spans the width of the box. The RT simulation used here was run at HealPix level $l=0$ for ray merging, corresponding to 12 independent directions per cell.  Ray noise can be seen as the graininess in this panel. {\it Middle:} The $\Gamma_{\rm HI}$ field from the same simulation used in the left panel, but averaged over 16 adjacent RT snapshots corresponding to a $\delta z \approx 0.05$ around $z=5.7$. {\it Right:} A version of the \textsc{$\dot{N}_{\rm ion} \propto L_{\rm UV}$} simulation run at HealPix level $l=1$, corresponding to 48 directions per cell.  Compared to the left panel, the ray noise is reduced in the middle and right panels, but large-scale features, e.g. peaks in the $\Gamma_{\rm HI}$ field and neutral island morphology, are broadly the same.}
\label{fig:gamma_noise}
\end{figure}

In this appendix we demonstrate that ray tracing discreteness noise does not affect our main conclusions on the opacity-density relation.  For the simulations in this paper, \textsc{FlexRT} has been configured to merge rays at HealPix level $l=0$, which corresponds to tracking the radiation field with $12$ unique directions in each cell. The finite number of rays leads inevitably to discreteness noise which can be seen as the graininess in the left panel of Figure \ref{fig:gamma_noise}. (The random rotations of the unit sphere performed at each time step translates into noise what would otherwise be spurious patterns in the RT fields.) There we show a cross section of the $z=5.7$ $\Gamma_{\rm HI}$ field in our \textsc{$\dot{N}_{\rm ion} \propto L_{\rm UV}$} model. The cross section is RT cell thick ($\Delta x = 1~\mathrm{cMpc}/h$), and spans the entire box in width. In \textsc{FlexRT}, the level of this noise depends on the HealPix level of the ray merging, as well as the number of rays per cell that are ``exempt'' from merging.  In our runs, the latter is $16$, such that the maximum number of rays per cell is $12 + 16 = 28$.   

To test the effect of ray tracing noise on our main results, we construct two other versions of the $\Gamma_{\rm HI}$ field: 1) we average $\Gamma_{\mathrm{HI}}$ from multiple RT steps over a time window shorter than the typical cell-crossing time of ionization fronts.  Specifically, we average over 16 RT time steps corresponding to $\delta z \approx 0.05$ around $z=5.7$.  Note that in this case we still merge at HealPix level $l=0$ (12 directions);  2) We have run version of the \textsc{$\dot{N}_{\rm ion} \propto L_{\rm UV}$} simulation that merges rays at HealPix level $l=1$, corresponding to 48 directions per cell.  These $\Gamma_{\rm HI}$ fields are shown in the middle and right panels of Figure \ref{fig:gamma_noise}, respectively.  In both cases, the graininess in the map is clearly reduced, but the large-scale features, e.g. intensity peaks and the morphology of neutral regions, remain largely the same.   

\begin{figure}
\begin{center}
\resizebox{0.6\linewidth}{!}{\includegraphics{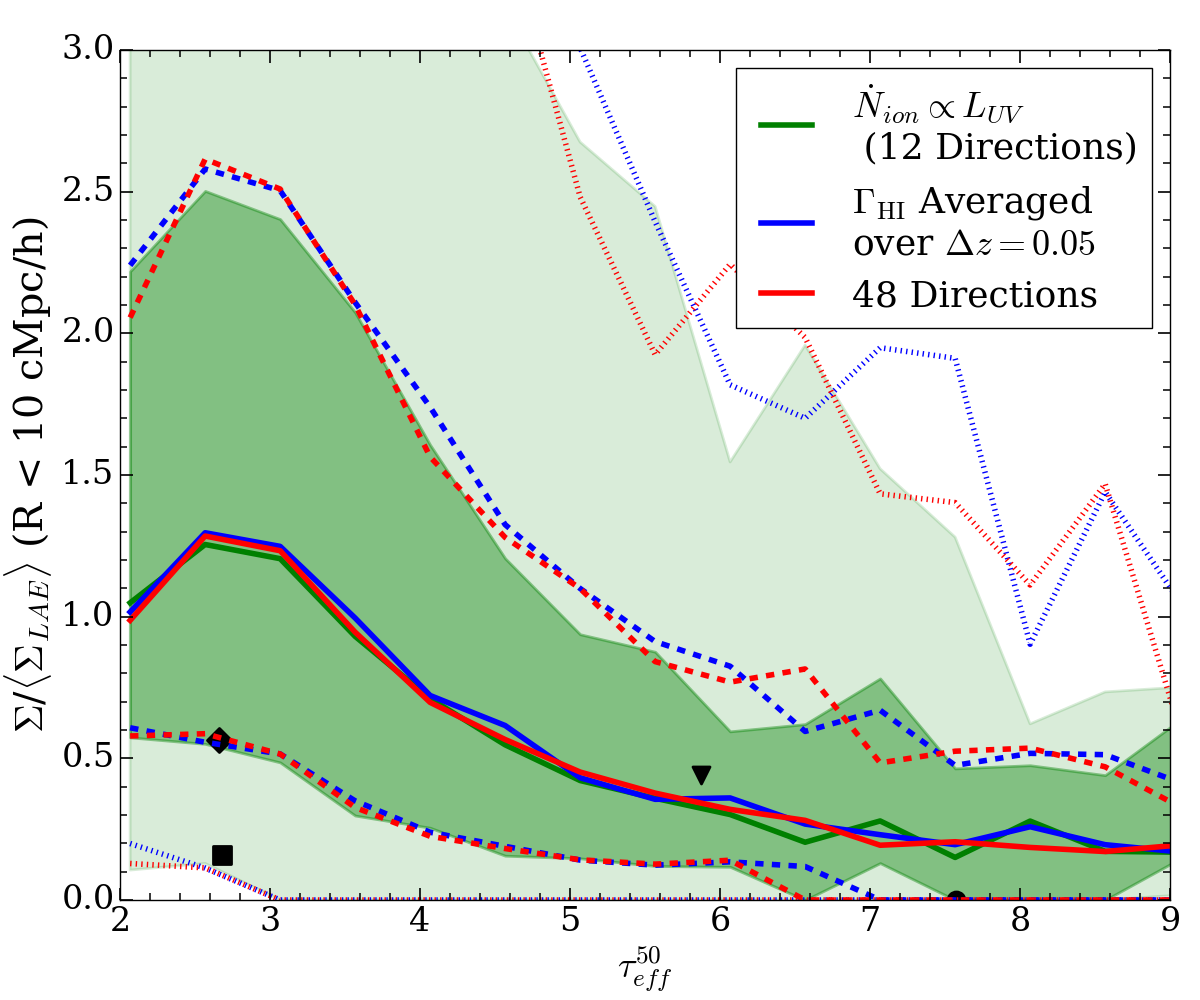}}
\end{center}
\caption{Effect of ray noise on the opacity-density relation. The opacity-density relation of the fiducial $\Nion \propto \LUV$ simulation (12 directions) is represented by the green curve and shaded regions, which show the median $\Sigma_{\rm LAE}/\langle \Sigma_{\rm LAE}\rangle$ versus $\tau_{\rm eff}^{50}$ and 68/98\% regions (c.f. Figure \ref{FIG:Sigma_vs_taueff}).   The blue and red sets of curves show the same for the two variations of the $\Gamma_{\rm HI}$ field in Figure \ref{fig:gamma_noise}.  The opacity-density relation is robust to our choice of merging rays at HealPix level $l=0$, which saves a significant amount of computational time.}
\label{fig:cg_SDvstau}
\end{figure}

Figure \ref{fig:cg_SDvstau} compares the derived opacity-density relations for the three variations depicted in Figure \ref{fig:gamma_noise}.  The green curves and contours correspond to the fiducial $\Nion \propto \LUV$ simulation. As before, the curve shows the median $\Sigma_{\rm LAE}/\langle \Sigma_{\rm LAE }\rangle$ versus $\tau^{50}_{\rm eff}$, while the dark and light shadings give the 68\% and 98\% regions, respectively. The blue and red sets of curves show the same for two variations introduced above.  We find that the fiducial simulation may under estimate the 68\% and 98\% upper limits by $\approx 5$ \%, but the opacity-density relations are otherwise quite similar. Importantly for the discussion in the main text, the probability of observing under-dense fields toward highly transmissive sight lines is not significantly affected by the ray noise.  We conclude that our main results are robust to our choice of merging rays at level $l=0$. It is worth noting that this choice (rather than $l=1$) greatly reduced the computational cost of the RT simulations run for this paper.  

\section{Effect of neutral island morphology on the opacity-density relation}
\label{appendix:neutralislandmorph}

\begin{figure*}
\begin{center}
\resizebox{0.6\columnwidth}{!}{\includegraphics{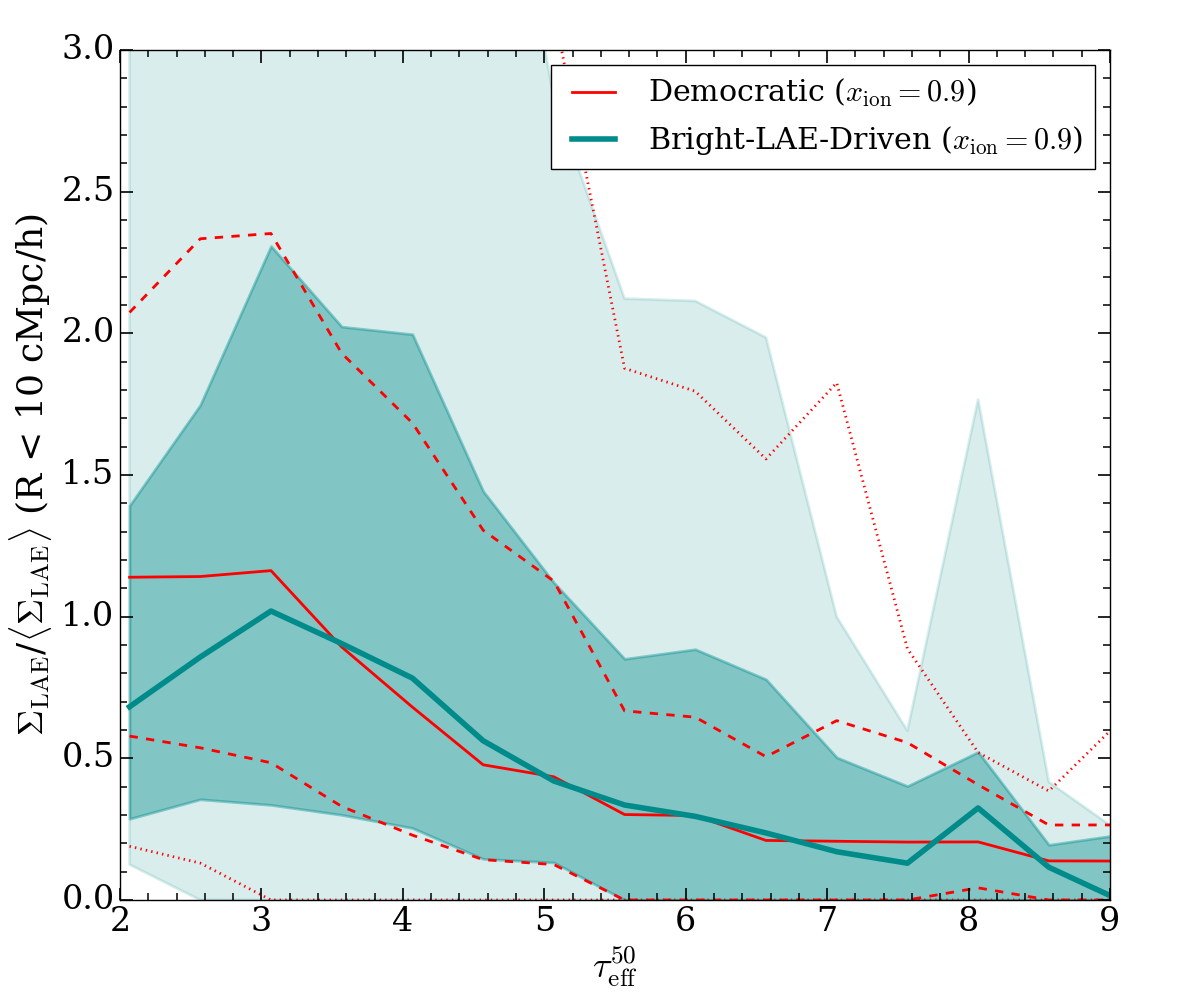}}
\end{center}
\caption{Effect of neutral island morphology on the opacity-density relation at fixed global neutral fraction of $x_{\rm HI} = 0.1$.  The red solid, dashed, and dotted curves show the median and 68/98\% regions from our fidicucial \textsc{Democratic} model.  The turquoise curve and shaded regions corresponds to a variation of our \textsc{bright-LAE-driven} model that has been re-scaled to have $x_{\rm HI} = 0.1$ at $z=5.7$ (see text for details). Owing to the more clustered sources, the neutral islands in this model are more clustered at the centers of the voids compared to the \textsc{Democratic} model at fixed neutral fraction.   This increases the physical cross section of hot, recently reionized gas in the voids, thereby increasing the likelihood that highly transmissive sight lines are associated with galaxy under-densities.}
\label{fig:source_clustering}
\end{figure*}

In this section we show that the insensitivity of the opacity-density relation to the source model that we found in \S \ref{sec:opac_dens} owes largely to the fact that all of our models have been calibrated to have the same mean Ly$\alpha$ forest transmission at $z\leq 6$. For models with a fixed neutral fraction at $z=5.7$, the morphology of neutral islands -- their sizes, shapes, and clustering properties -- has an important effect on the opacity-density relation.  Generally speaking, neutral islands tend to be larger and more clustered when the ionizing photon budget is dominated by more biased sources \cite{Cain2023}, and this can have a particularly strong affect on the low-$\tau_{\rm eff}^{50}$ end of the opacity-density relation. 

To illustrate this, we compare two models with the same global neutral fraction at $z=5.7$, but with very different source populations.  For one case we use our \textsc{Democratic} model from \S \ref{sec:opac_dens}, which has $x_{\rm HI}(z=5.7) = 0.1$. For the other model, we re-scale the $z=5.4$ snapshot from the \textsc{Bright-LAE-driven} simulation, which also has $x_{\rm HI} = 0.1$.  Specifically, we apply the neutral island, temperature, and $\Gamma_{\rm HI}$ fields from this later snapshot to construct our mock Ly$\alpha$ forest segments, re-scaling $\Gamma_{\rm HI}$ in ionized regions to match the mean forest transmission at $z=5.7$.  This approximates a simulation in which the tail end of reionization proceeds more rapidly than in our fiducial \textsc{Bright-LAE-driven} model, such that the $x_{\rm HI}$ is lower at $z=5.7$.

Figure \ref{fig:source_clustering} compares the opacity-density relations between these two test models.  The red curves show the median and 68/98\% limits from the \textsc{Democratic} model, while the turquoise curve and shaded regions correspond to the re-scaled \textsc{Bright-LAE-driven} model. We emphasize that this comparison is at fixed $x_{\rm HI} = 0.1$.  Here we see important differences on the low-$\tau^{50}_{\rm eff}$ end of the relation. Under-densities around highly transmissive sight lines are more likely in the \textsc{Bright-LAE-driven} model. Indeed, the probability of observing $\leq 1$ LAE within a radius of $R = 10~\mathrm{cMpc}/h$ is 10\% in that model, compared to the 4\% probability in the \textsc{Democratic} model. The increased probability in the former occurs because the neutral islands are more clustered, i.e. they are less spread out throughout the voids compared to the \textsc{Democratic} model.  Hence, even at fixed global neutral fraction,  the likelihood of intersecting hot, recently reionized gas in the voids is higher in the \textsc{Bright-LAE-driven} model.  This highlights the role that neutral island morphology plays in setting the frequency of sight lines like J359.

\bibliographystyle{JHEP}
\bibliography{JCAP_Folder/jcap_references}

\end{document}